# Unsupervised detection and classification of heartbeats using the dissimilarity matrix in PCG signals

J. Torre-Cruz *[a], D. Martinez-Muñoz[a], N. Ruiz-Reyes[a], A.J. Muñoz-Montoro[b], M. Puentes-Chiachio[c], F.J. Canadas-Quesada[a]

[a]*Department of Telecommunication Engineering. University of Jaen, Campus Cientifico-Tecnologico de Linares, Avda. de la Universidad, s/n, 23700 Linares, Jaen, Spain*
[b]*Department of Computer Science, University of Oviedo, Campus de Gijón, s/n, Gijón, 33203, Spain*
[c]*Cardiology, University Hospital of Jaen, Av. del Ejercito Espanol, 10, 23007 Jaen, Spain*

## Abstract

*Background and objective*: Auscultation is the first technique applied to the early diagnose of any cardiovascular disease (CVD) in rural areas and poor-resources countries because of its low cost and non-invasiveness. However, it highly depends on the physician's expertise to recognize specific heart sounds heard through the stethoscope. The analysis of phonocardiogram (PCG) signals attempts to segment each cardiac cycle into the four cardiac states (S1, systole, S2 and diastole) in order to develop automatic systems applied to an efficient and reliable detection and classification of heartbeats. In this work, we propose an unsupervised approach, based on time-frequency characteristics shown by cardiac sounds, to detect and classify heartbeats S1 and S2. *Methods*: The proposed system consists of a two-stage cascade. The first stage performs a rough heartbeat detection while the second stage refines the previous one, improving the temporal localization and also classifying the heartbeats into types S1 and S2. The first contribution is a novel approach that combines the dissimilarity matrix with the frame-level spectral divergence to locate heartbeats using the repetitiveness shown by the heart sounds and the temporal relationships between the intervals defined by the events S1/S2 and non-S1/S2 (systole and diastole). The second contribution is a verification-correction-classification process based on a sliding window that allows the preservation of the temporal structure of the cardiac cycle in order to be applied in the heart sound classification. The proposed method has been assessed using the open access databases PASCAL, CirCor DigiScope Phonocardiogram and an additional sound mixing procedure considering both Additive White Gaussian Noise (AWGN) and different kinds of clinical ambient noises from a commercial database. *Results*: The proposed method outperforms the detection and classification performance of other recent state-of-the-art methods. Although our proposal achieves the best average accuracy for PCG signals without cardiac abnormalities, 99.4% in heartbeat detection and 97.2% in heartbeat classification, its worst average accuracy is always above 92% for PCG signals with cardiac abnormalities, signifying an improvement in heartbeat detection/classification above 10% compared to the other state-of-the-art methods evaluated. *Conclusions*: The proposed method provides the best detection/classification performance in realistic scenarios where the presence of cardiac anomalies as well as different types of clinical environmental noises are active in the PCG signal. Of note, the promising modelling of the temporal structures of the heart provided by the dissimilarity matrix together with the frame-level spectral divergence, as well as the removal of a significant number of spurious heart events and recovery of missing heart events, both corrected by the proposed verification-correction-classification algorithm, suggest that our proposal is a successful tool to be applied in heart segmentation.

*Corresponding author. Tlf.: (+34) 953648592
*Email address:* jtorre@ujaen.es (J. Torre-Cruz *)

## 1. Introduction

The World Health Organization (WHO) reports that cardiovascular diseases (CVDs), such as heart failure, are among the leading causes of death worldwide, claiming approximately 18 million lives each year [1]. In Spain, more than 125,000 people die each year from CVDs, with around 315 acute myocardial infarctions per 100,000 male inhabitants and 80 per 100,000 female inhabitants, with an annual increase of 2% in the occurrence of non-fatal infarctions within the male population [2]. Due to the seriousness of this situation, the European Commission is building the Horizon Europe's Strategic Plan 2021-2024 in order to achieve an improved evidence-based health policies, and more effective solutions for health promotion and disease prevention [3]. Thus, most research efforts of the biomedical signal processing community are being focused on early identifying people at increased risk of developing CVDs in order to accelerate the application of medical treatment to prevent premature death, since WHO specifies that 80% of both premature heart diseases and strokes are preventable [4].

From a physiological standpoint, a normal heart rate (HR) can be defined as a repetitive sequence of two primary heart sounds, S1 (known as lub) and S2 (known as dub), during each heart cycle [5]. The sounds S1 and S2 follow specific temporal-spectral patterns, given that the HR associated to an adult person ranges between 60-80 beats per minute (bpm) and can be considered a predictor of cardiovascular risk whenever a resting HR is above 80 bpm [6]. In fact, both the HR and the spectral content associated to heart sounds can reveal useful medical information on cardiovascular risk, being the continuous HR monitoring most important for the early diagnosis of CVDs [7]. It is well known that S1 occurs at systole phase, closing the mitral and tricuspid valves, while S2 occurs at diastole phase, closing the aortic and pulmonic valves. Although most of the time it is assumed that systole is shorter than diastole [8, 9, 10], the duration of systole is commonly considered relatively constant [11, 12, 13] whereas the length of diastole varies according to the HR [14, 12]. Comparing S1 and S2, S1 usually shows a low-pitch (about 20-150 Hz) with an average longer duration (about 70-160 ms) and higher amplitude. However, S2 often shows a high-pitch (about 20-200 Hz) sound with an average shorter duration (about 60-140 ms) and lower amplitude [15, 16, 13]. In unhealthy adults, in addition to the primary heart sounds S1 and S2, there are also abnormal heart sounds such as murmurs characterised by the concentration of most of their energy in the spectral range of 15-700 Hz and normally located within the systole or diastole [13, 16].

Techniques such as echocardiography, electrocardiography (ECG) or medical resonance imaging (MRI) are usually used to diagnose CVDs, however, these techniques involve a higher cost compared to auscultation, so it is difficult for these techniques to be applied in low- and middle-income countries as it is not an economically viable option. Alternatively, auscultation can diagnose effectively CVDs [17] and it remains the most commonly applied technique in poor-resources countries to early diagnose any CVD because of its low cost, non-invasiveness and user-friendliness. Nevertheless, the reliability of this technique highly depends on the human auditory system, the sound quality captured by the stethoscope and the physician's expertise in order to recognize and interpret the meaning of the sounds heard through the medical devices [17, 18]. As a consequence, the analysis of heart sounds by phonocardiography (PCG) has become a research topic of high interest in the biomedical signal processing community in recent years, since a PCG signal is a recording of heart sounds through the cardiac cycles by means of a digital stethoscope or microphone mainly located over the chest wall of the patient [15], as shown in Figure 1. PCG signal analysis, in addition to providing the HR and duration of systole and diastole as does ECG, can extract additional information (e.g. energy distribution and spectral envelope of heart sounds in the time-frequency domains) of high interest to medical professionals and engineers, because it is best suited to further analyse both normal and abnormal heart sounds.

Focusing on PCG signals, the first task is usually the HR estimation, however, it is highly associated with heart segmentation [20]. The aim of the heart segmentation (HS) in the PCG signal is twofold: i) localization of frames in which the heart sounds are active and, ii) classification of the previous cardiac frames into S1 and S2 [5, 7]. As a result, the four states of each heart cycle are defined: S1, S2, systole (time interval between S1 and S2), diastole (time interval between S2 and S1) [21] and the HR, implicitly. Therefore, the accurate temporal localization of this sequence of repetitive S1 and S2 heart sounds is the key [22, 17] to developing any heart sound detection [23, 24, 7] [25, 26] or classification system [7, 25, 26, 27, 28, 29], since their performances highly depend on the accuracy obtained in the segmentation stage, in order to extract and classify specific time-frequency features.

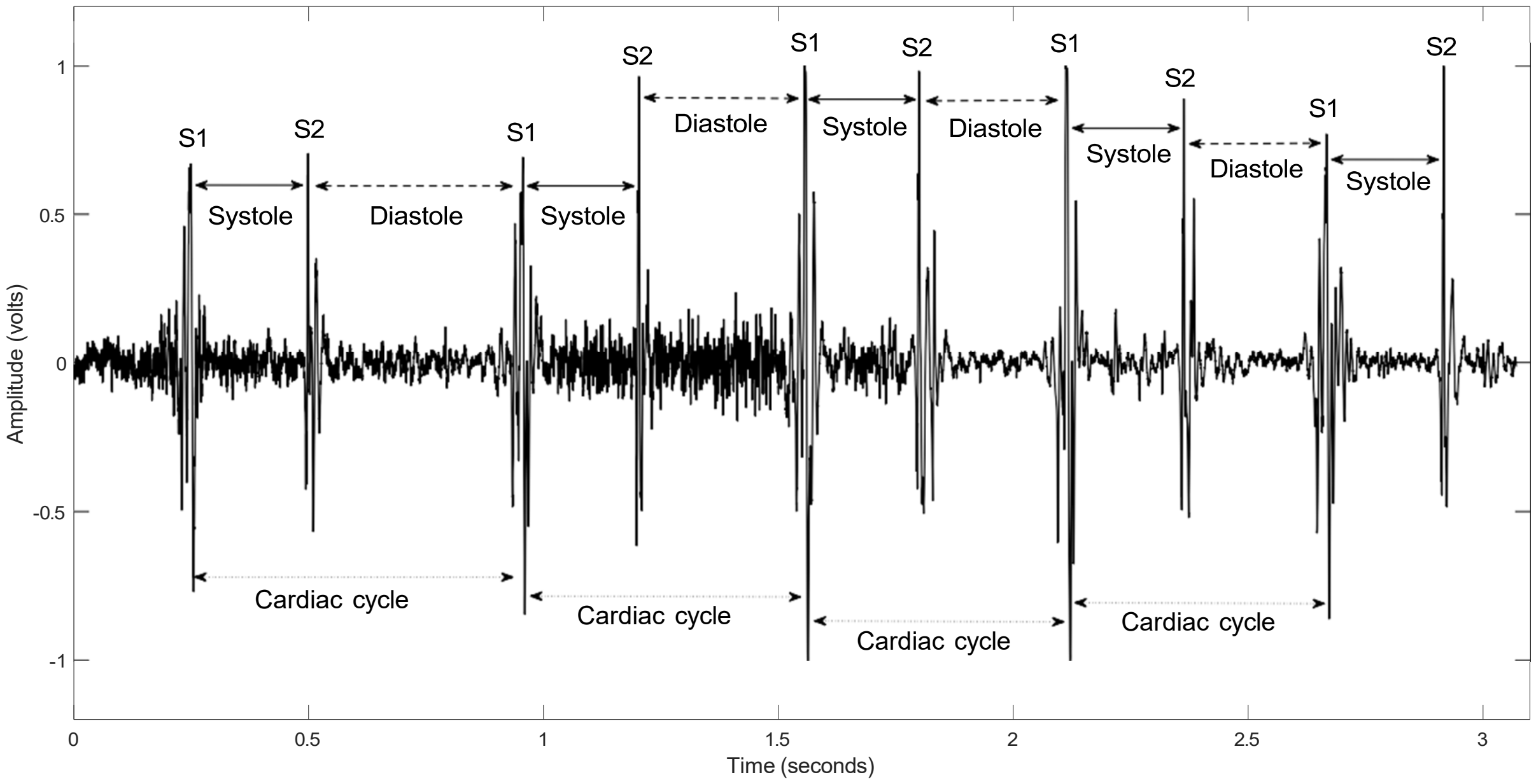

Figure 1: An example of PCG signal, obtained from database PASCAL (173 1307973611151LC.wav) [19], with annotations of the parts that compose each cardiac cycle. It can be observed that the duration of systole is approximately the same in each cardiac cycle but this is not the case when the diastole is analysed.

Focusing on the temporal repeatability shown by specific sounds in the real-world, temporal regularity plays a facilitating role in pattern detection according to [30, 31]. The human auditory system can easily isolate those sounds by identifying them as repeating patterns active in the input signal as occurs with the soloist and rhythmic accompaniment in a musical sound context. In the last years, many signal processing approaches have been proposed in order to exploit the temporal repeatability of sounds in different research fields, such as audio [32, 33, 34], image [35] and biomedicine [36, 18]. Specifically, the concept of self-similarity was initially proposed to visualise the time structure of music [37] and later to extract the repeating background from the foreground in musical signals but authors aimed that it could be generalised to any kind of repeating patterns because the similarity information showed a high robustness to successfully extract intermittent or variable periods from a repeating structure [33]. With this in mind, it would be interesting to explore the event detection and classification applied to heart sounds using the previously mentioned similarity information since the heart sounds are widely considered the most intuitive and repetitive pattern in the human nature [38] [39].

In literature, numerous heart segmentation works have been proposed in the last decades, being the most common approaches categorised in Fourier transform [40, 41], envelope-based [42, 43], wavelet [44, 45, 46], energy-based [47, 38, 48, 23, 49], S-transform [43, 39], Support vector machine (SVM) [50, 51], Spectral clustering [25], Hidden markov model (HMM) [52, 53], empirical mode decomposition [54, 55], Homomorphic filtering [56, 7], Neural networks [22, 17, 9, 57] and Mel-frequency cepstral coefficients (MFCC) [47, 26]. Authors [46] detailed a novel method for detecting and classifying heart sounds and murmurs by means of empirical wavelet transform-based PCG signal decomposition, the Shannon entropy envelope extraction, the instantaneous phase-based boundary determination, heart sound and murmur parameter extraction, the systole/diastole discrimination and the decision rules-based murmur classification. Mubarak et al. [7] proposed the concept of quality assessment before localization, feature extraction and classification of heart sounds. Specifically, the signal quality is assessed by predefined criteria based upon the number of peaks and zero crossing of the PCG signal. Once quality assessment is performed, then heart beats within the PCG signal are localised, which is done by envelope extraction using homomorphic envelogram and finding prominent peaks. In order to classify localised peaks into S1 and S2, temporal and time-frequency based statistical features have been used. Support Vector Machine using radial basis function kernel is used for classification of heart beats into S1 and S2 based upon extracted features. In [9], a convolutional network architecture, inspired by a network used for image segmentation, was combined with a sequential temporal modeling procedure in order to provide segmentation results compatible with the sequential nature of heart sounds. Convolutional Neural Network (CNNs) were also incorporated in segmentation algorithms using HMMs. Results showed that CNNs improved performance in recovering the exact positions of fundamental heart sounds in PCGs. According to [25], an unsupervised approach to detect the positions of heart sound events S1 and S2 in PCGs recordings is presented. Gammatone filter bank features and a spectral clustering technique were used for the segmentation of S1/S2 and non-S1/S2 heart sound events. In [57], a real-time algorithm for heart segmentation based on CNN is developed testing feature extraction methods using the spectrum, envelopes, and 1D-CNN. Moreover, this approach is combined with the Viterbi algorithm to correct the heart sound state order errors in the point-by-point classification algorithm. In [26], a complete algorithm for automatic heartbeat detection and disorder discrimination is presented separating S1 and S2 by means of a power threshold, being the frame duration dynamically estimated according to the duration of the sound S1 or S2. Lastly, two methods to combine MFCC estimates are proposed.

In this paper, we aim to develop an unsupervised detection and classification of S1 and S2 heartbeat sounds combining the frame-level spectral divergence from the dissimilarity matrix applied to PCG recordings. In a first stage, an initial heart sound detection is estimated assuming both the well known repeatability shown by the heartbeats and the additional temporal constraints among S1, S2, systole and diastole. In a second stage, a more accurate detection is achieved improving the previous estimation and the classification between S1 and S2 heartbeats is obtained using temporal assumptions between the systole and diastole.

The rest of the paper is organised as follows: the proposed method for detecting and classifying S1 and S2 heartbeat sounds is detailed in Section 2. Section 3 describes the dataset, metrics, setup, the state-of-the-art methods and results. The performance of the proposed method is discussed in Section 4. Finally, conclusions and future work are presented in Section 5.

## 2. Proposed method

From the PCG signal $x[m]$ captured by a digital device (e.g., stethoscope or microphone), the goal of the proposed method is to detect and classify the heart sounds S1 and S2. In this respect, we propose an unsupervised approach based on the time-frequency characteristics of heartbeats to model the typical behaviour of the heart sounds S1 and S2. As can be observed in Figure 2, the proposed method is composed of two stages: Rough heartbeat detection (Stage I) and Fine heartbeat detection (Stage II). Briefly, Stage I performs a preliminary heartbeat detection based on temporal repeatability of the heart sounds S1 and S2 and the temporal relationships among heartbeats (S1 and S2) and the intervals between them (systole and diastole). On the other hand, Stage II refines the temporal localization of the heartbeats detected in the previous stage and classifies them into S1 or S2, assuming temporal relationships between systole and diastole.

### *2.1. Stage I. Rough heartbeat detection*

The objective of this stage is to obtain an approximate detection of the heartbeats S1 and S2, assuming the following two common temporal-spectral behaviours observed in most heart sounds: i) heart sounds can be modelled as spectral structures that repeat over time, showing similar frequency patterns between S1 and S2 but distinct from those patterns that can be found in systole and diastole [25]; and ii) the duration of both S1 and S2 is shorter than systole and diastole [15]. Based on the two cardiac behaviors mentioned above, we propose the dissimilarity matrix, a novel modification of the standard similarity matrix [33], that exploits the above typical behaviors shown by most heart sounds. Specifically, this stage is divided into three steps which are detailed below.

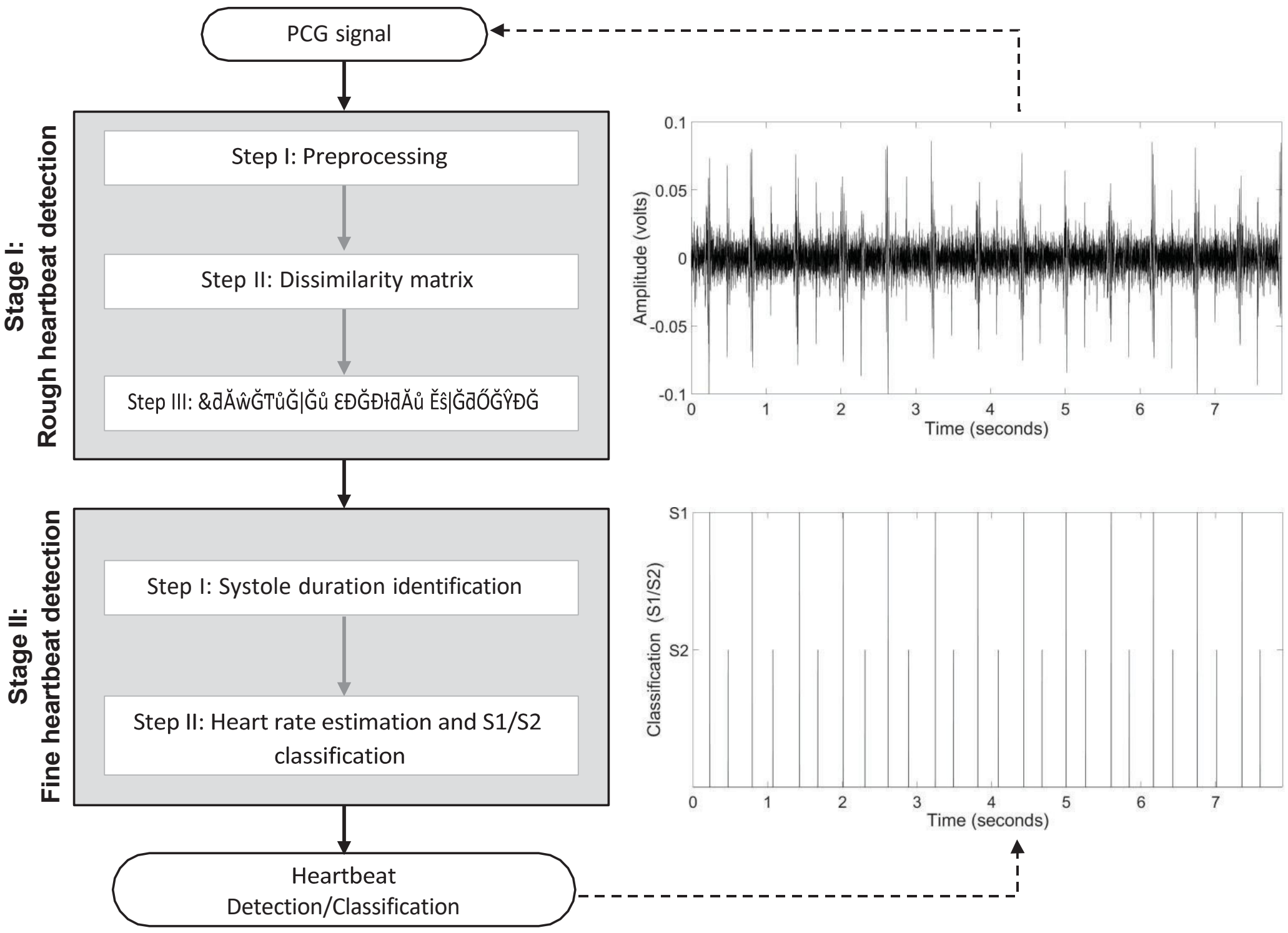

Figure 2: Block diagram of the proposed method.

#### 2.1.1. Step I: Preprocessing

The magnitude spectrogram $\mathbf{X}$ associated with the input PCG signal $x[m]$ is obtained applying the Short-Time Fourier Transform (STFT) with a Hamming windows of $N$ samples and $S$, in percentage, overlap in order to analyse the spectral content shown by heart sounds. Each magnitude spectrogram $\mathbf{X}$ is composed of $T$ frames, $F$ frequency bins and a set of time-frequency units $X_{f,t}$, being $f = 1, ..., F$ and $t = 1, ..., T$. Specifically, each unit $X_{f,t}$ represents the $f^{th}$ frequency bin at the $t^{th}$ frame. Moreover, $\mathbf{X}$ is normalised to achieve independence both in size and scale and as result its energy equals 1.0. Thus, the normalised magnitude spectrogram $\overline{\mathbf{X}}$ is computed in Equation (1).

$$\overline{\mathbf{X}} = \frac{\mathbf{X}}{\sqrt{\sum_{f=1}^{F} \sum_{t=1}^{T} X_{f,t}^2}} \tag{1}$$

Next, a band-pass filter with cut-off frequencies from 20 to 200 Hz is applied to the normalised magnitude spectrogram $\overline{\mathbf{X}}$ to focus the analysis on the spectral structures of the heart sounds since it is well known that most of the energy of the sounds S1 and S2 is concentrated within the bandwidth 20 Hz to 200 Hz [15, 16, 13]. Note that to avoid complex nomenclature throughout the paper, the variable $\mathbf{X}$ is hereinafter referred to the filtered version of the previous normalised magnitude spectrogram $\overline{\mathbf{X}}$. Figure 3 shows an example of PCG signal and its filtered spectrogram. It can be observed that the heart sounds show a repetitive structure over time, showing significant similarity among their spectral patterns across all cardiac cycles. Specifically, the spectral patterns related to S1 and S2 share similar time-frequency content, as well as the spectral patterns of beat-to-beat separation intervals such as systole and diastole. Moreover, the duration of S1 and S2 is always shorter compared to systole and diastole. It should be indicated to the reader that the recording analyzed in Figure 3 will be used throughout section 2 to explain the procedure applied for each step of the proposed method.

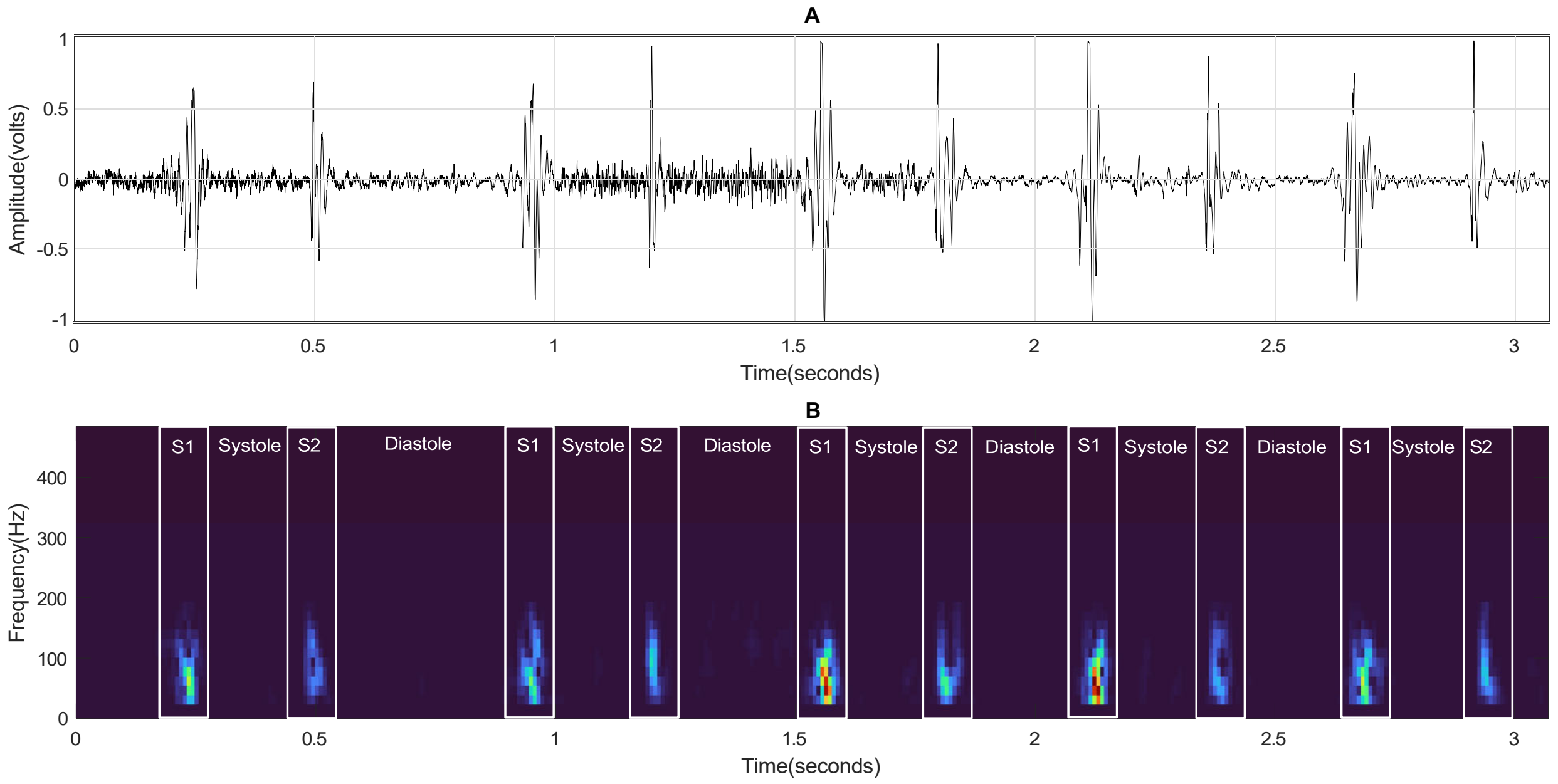

Figure 3: Example of recording ”173_1307973611151_C.wav” obtained from database PASCAL [19]. A) Input PCG signal $x[m]$. B) Filtered spectrogram $\mathbf{X}$ in which the annotations of the four states (S1, S2, systole and diastole) that compose each cardiac cycle are shown.

### *2.1.2. Step II: Dissimilarity matrix*

In order to extract the repetitive structure shown by heart sounds from the PCG spectrogram $\mathbf{X}$, and thus locate in time the occurrence of the patterns S1 and S2, we propose the dissimilarity matrix $\mathbf{D}$ as a new version of the standard similarity matrix [33] particularly adapted to heart sounds.

The dissimilarity matrix $\mathbf{D}$ is a two-dimensional representation where each point $D_{i,j}$ measures the spectral dissimilarity between two frames $X_i$ and $X_j$ of the input PCG spectrogram $\mathbf{X}$, where $i = 1, ..., T$ and $j = 1, ..., T$. In other words, the higher the dissimilarity between the spectral content of two frames, the higher the dissimilarity value and vice versa. The dissimilarity matrix obtains their minimum values in the main diagonal $D_{i,i}$ because each $i^{th}$ frame is always the most similar to itself at any time. In general terms, the row or column $i^{th}$ belonging to the matrix $\mathbf{D}$ provides the degree of dissimilarity of the $i^{th}$ frame with all frames that compose the spectrogram $\mathbf{X}$. In this paper, we propose to use the Kullback-Leibler divergence $d_{KL}(X_i|X_j)$ to provide the degree of dissimilarity between each frame pair $(X_i|X_j)$ of the spectrogram $\mathbf{X}$ since $d_{KL}(X_i|X_j)$ provides a scale-invariant dissimilarity, and as a result, low energy sound components of $X_i$ and $X_j$ bear the same relative importance as high energy ones. The calculation of the dissimilarity matrix $\mathbf{D}$ is shown in Equation (2).

$$D_{i,j} = d_{KL}(X_i|X_j) = \sum_{f=1}^{F} X_{f,i} log \frac{X_{f,i}}{X_{f,j}} - X_{f,i} + X_{f,j} \qquad \forall i,j \in [1,T] \tag{2}$$

In this paper, an event S1/S2 means a heartbeat sound of type S1 or S2, but we do not know to which of the two it belongs. As previously mentioned, the temporal information provided by the dissimilarity matrix $\mathbf{D}$ could be useful to discriminate between spectral content belonging to heartbeats S1/S2 and beat-to-beat separation intervals, that is, diastole/systole. More specifically, those frames S1/S2 show a high degree of dissimilarity compared to those ones associated to systole/diastole, but a low degree of dissimilarity compared to other frames S1/S2 related to other heartbeats located at any time. Likewise, frames systole/diastole report a high degree of dissimilarity with frames S1/S2, but a low degree of dissimilarity compared to other frames belonging to systole/diastole. These facts can be observed in Figure 4.

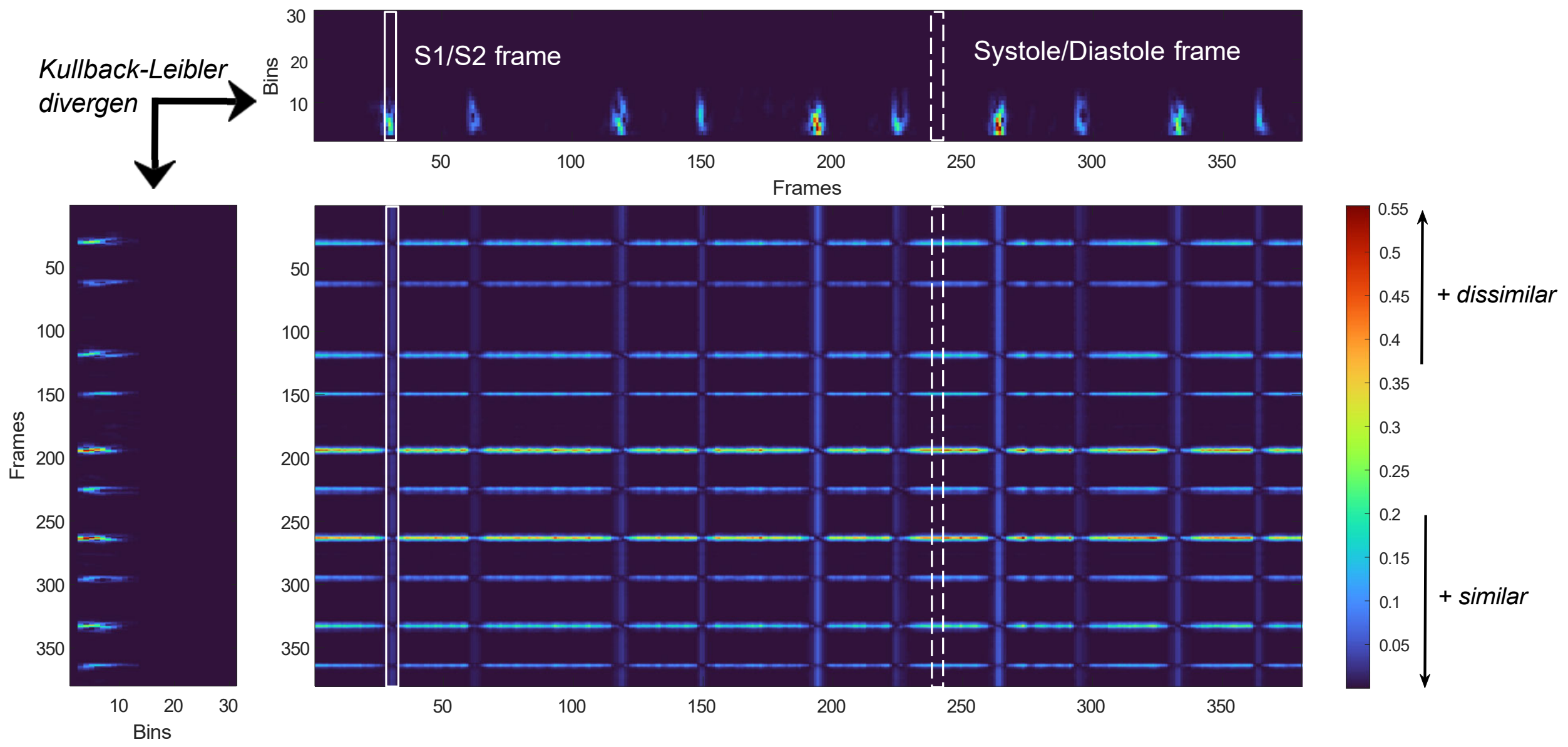

Figure 4: Example of recording ”173 1307973611151 C.wav” from the database PASCAL [19]. The top and left subfigures show the filtered spectrogram **X** of the input PCG signal. The middle subfigure shows the dissimilarity matrix **D**, where each point $D_{i,j}$ measures the Kullback-Leibler divergence $d_{KL}(X_i|X_j)$ between two frames $X_i$ and $X_j$ of the spectrogram **X**. The continuous white rectangle marks column 36 of matrix **D**, showing the values of the Kullback Leibler divergence between the frame $X_{36}$ (frame S1/S2) and each frame of spectrogram **X**, i.e., $d_{KL}(X_{36}|X_j)$ where $j = 1, ..., T$. The dashed white rectangle marks column 245 of matrix **D**, showing the values of the Kullback Leibler divergence between the frame $X_{245}$ (systole/diastole frame) and each frame of spectrogram **X**, i.e., $d_{KL}(X_{245}|X_j)$ where $j = 1, ..., T$. As shown, the frames S1/S2 show a high degree of dissimilarity compared to those ones located within the intervals systole/diastole and vice versa.

#### 2.1.3. Step III: Frame-level spectral divergence

We assume that the temporal localization of the frames S1/S2 could be based on the fact that sounds S1/S2 are shorter compared to sounds active within systole/diastole [15]. Therefore, most of the time the input PCG signal $x[m]$ is composed of the states systole/diastole with a smaller proportion of intervals S1/S2. As can be seen in Figure 5A, the dissimilarity matrix **D** can model the frames S1/S2 by means of lobes with high temporal width, because these lobes represent the dissimilarity regarding to long intervals such as systole or diastole (see Figure 5B). However, the dissimilarity matrix **D** models the frames associated to systole or diastole as lobes with a very low temporal width because these lobes represent the dissimilarity regarding to short intervals such as S1 or S2 (see Figure 5C). As a consequence, we propose to calculate the frame-level spectral divergence $\alpha_i$, that is, the sum of the spectral divergences associated with each $i^{th}$ frame with respect to all other frames of the matrix **D** since this technique provides a discriminative measure to classify S1/S2 and systole/diastole frames. In other words, each value $\alpha_i$, approximated by the integration over an interval by decomposing the area into trapezoids with more easily computable areas [58], represents the total area $\alpha_i$ under the curve defined by the dissimilarity lobes from each row $i^{th}$ of the dissimilarity matrix **D**. Figure 6 shows the frame-level spectral divergence $\alpha_i$ from the dissimilarity matrix **D**, clearly discriminating between frames S1/S2 and frames systole/diastole. Finally, a standard peak detector is applied on the vector $\alpha$ to obtain the frames $\lambda_k$, $k = [1, ..., K]$, where $K$ is the number of sounds S1/S2 estimated. These frames $\lambda_k$ are those ones with the highest frame-level spectral divergence, so they are labelled as frames S1/S2 because heart sounds S1 or S2 are active within them.

The pseudo-code for Stage I to roughly estimate the frames S1/S2 $\lambda_k$ (at the frame level) by combining the dissimilarity matrix **D** with the frame-level spectral divergence $\alpha_i$ is detailed in Algorithm 1.

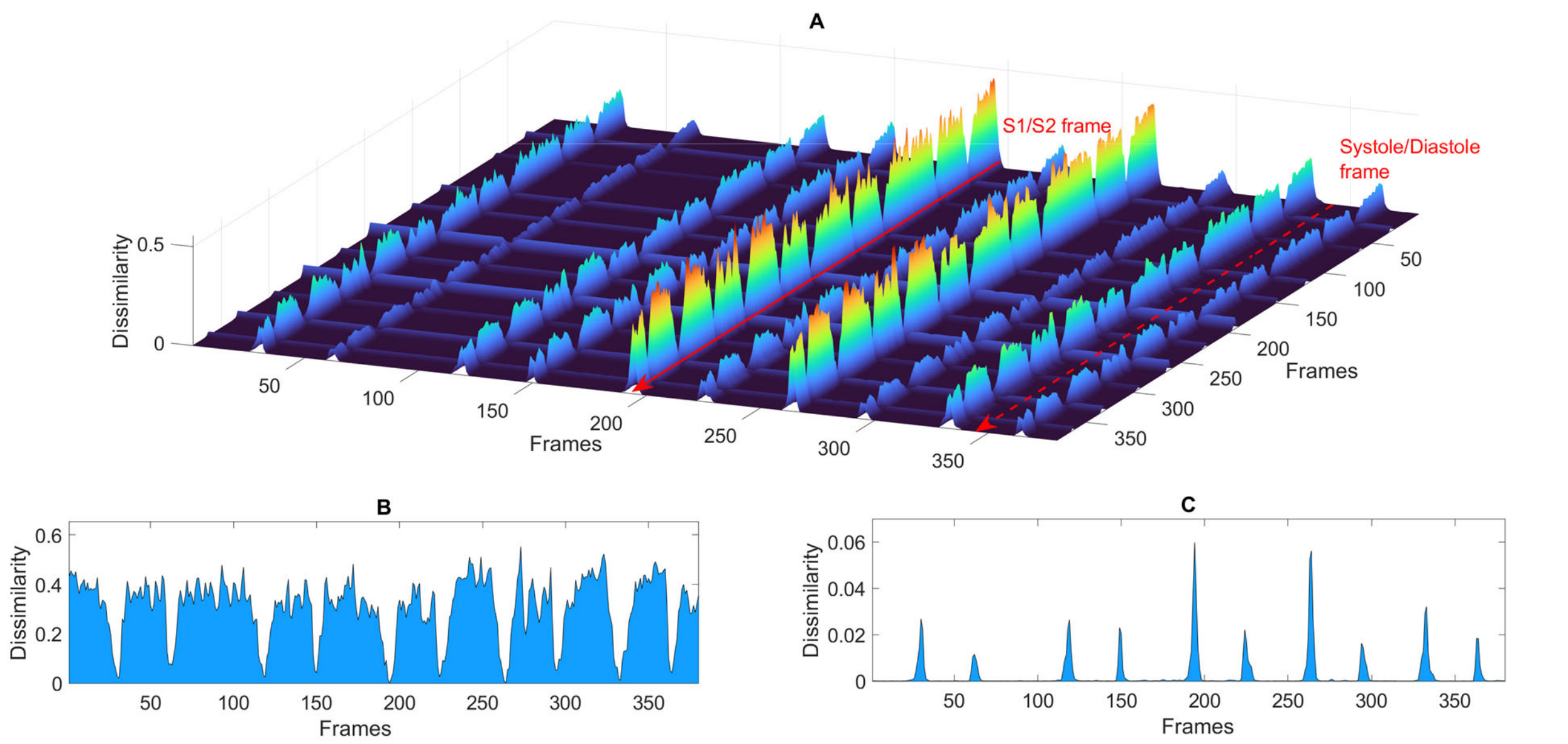

Figure 5: Example of recording ”173_1307973611151_C.wav” obtained from database PASCAL [19]. A) 3D representation of the dissimilarity matrix **D**. B) 2D representation of the area $a_{192}$ belonging to the frame S1/S2 obtained from **D** (solid red arrow). C) 2D representation of the area $a_{347}$ of a frame systole/diastole obtained from **D** (dashed red arrow).

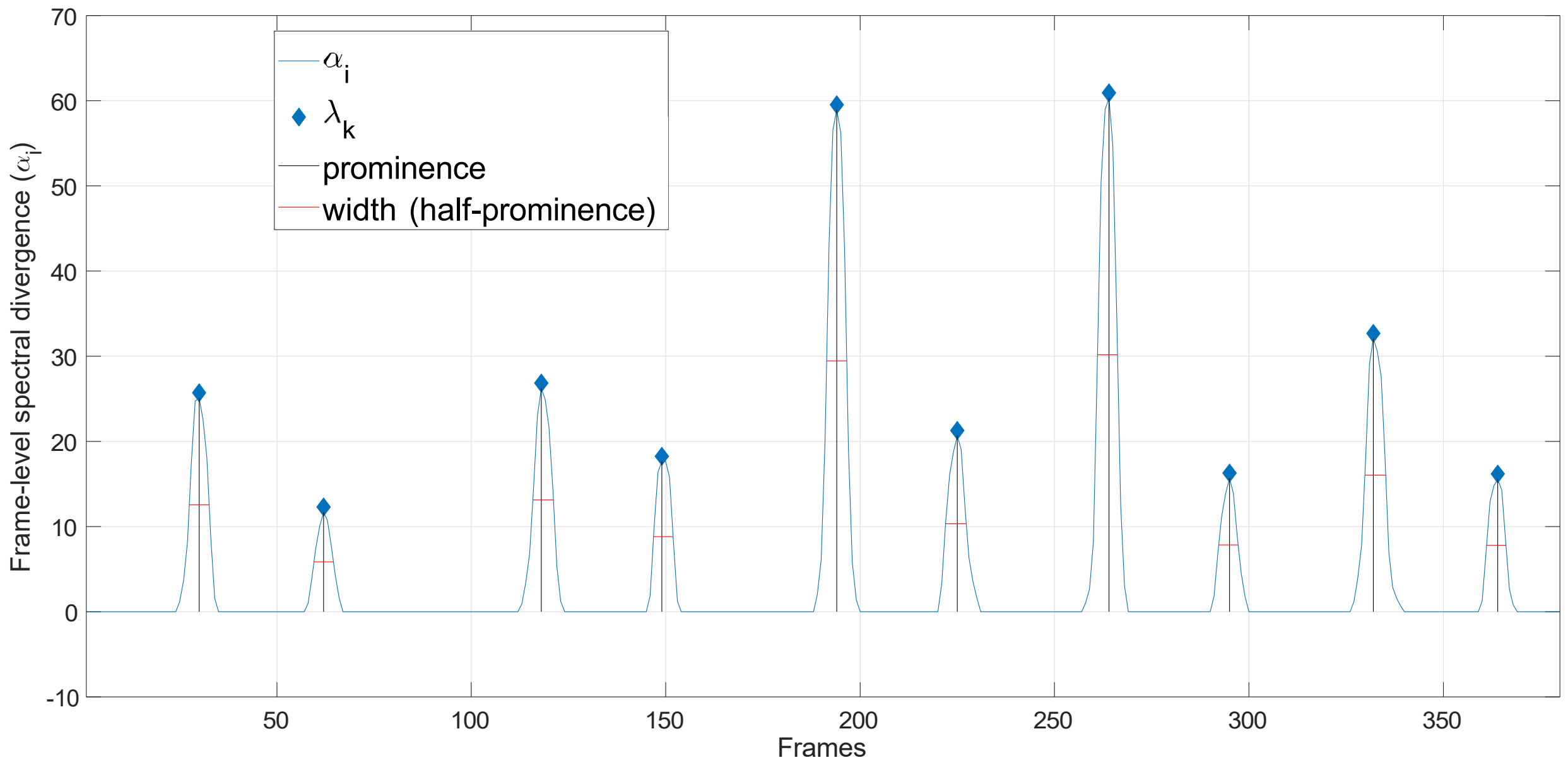

Figure 6: Example of recording ”173_1307973611151_C.wav” obtained from database PASCAL [19]. The frame-level spectral divergence $\alpha_i$ for each row (frame) $i = 1, ..., T$ of matrix **D**, obtaining the detection of heart sounds S1 and S2 at frame level $\lambda_k$.

### *2.2. Stage II. Fine heartbeat detection*

The objective of this stage is to refine the temporal localization of the frames S1/S2 $\lambda_k$ obtained in Stage I, and subsequently classify them as S1 or S2. As previously mentioned, we assume that the temporal duration of cardiac systole tends to be constant over time compared to diastole intervals, so HR changes mostly due to the temporal variation of the cardiac diastole over time. Based on this, we propose a method

**Algorithm 1** Rough heartbeat detection

**Require**: $x[m]$, $N$ and $S$.

**Step I: Preprocessing**

Compute the magnitude PCG spectrogram $\mathbf{X}$ by means of the STFT using a Hamming windows of $N$ samples with $S$ overlap.

Compute the normalised spectrogram $\overline{\mathbf{X}}$ using Equation (1).

Apply a band-pass filter with cut-off frequencies from 20 to 200 Hz.

**Step II: Dissimilarity matrix**

Calculate the dissimilarity matrix $\mathbf{D}$ using Equation (2).

**Step III: Frame-level spectral divergence**

Calculate the frame-level spectral divergence $\alpha_i$ for each row $i^{th}$ of the matrix $\mathbf{D}$.

Localization of the S1/S2 frames, denoted as $\lambda_k$, from the frame-level spectral divergence $\alpha_i$.

**return** $\lambda_k$

based on a sliding window algorithm that validates and corrects the previously cardiac frames $\lambda_k$ and then, classifies them as S1 or S2. Specifically, this stage is divided into two steps which are detailed below.

### *2.2.1. Step I: Systole duration identification*

The cardiac frames $\lambda_k$ are converted to cardiac samples $\delta_k$ in order to improve the accuracy of the events S1/S2 detected in the previous stage. For this purpose, we propose to use a window of size $N$ along the input PCG signal $x[m]$ to locate the sample $\delta_k$ associated to each frame $\lambda_k$. To do that, we select the set of samples $\rho_k$ belonging to the cardiac frame $\lambda_k$ as follows,

$$\rho_k = [\lfloor \lambda_k \cdot S \cdot N \rfloor + 1, ..., \lfloor \lambda_k \cdot S \cdot N \rfloor + N] \qquad (3)$$

where $\lfloor \ \rfloor$ denotes the rounding operator to the nearest integer value to zero. Next, we determine that a sound S1/S2 appears by looking for the sample $\delta_k$ with the largest amplitude within the set of samples $\rho_k$. The procedure how the sample $\delta_k$ is computed can be observed in Figure 7.

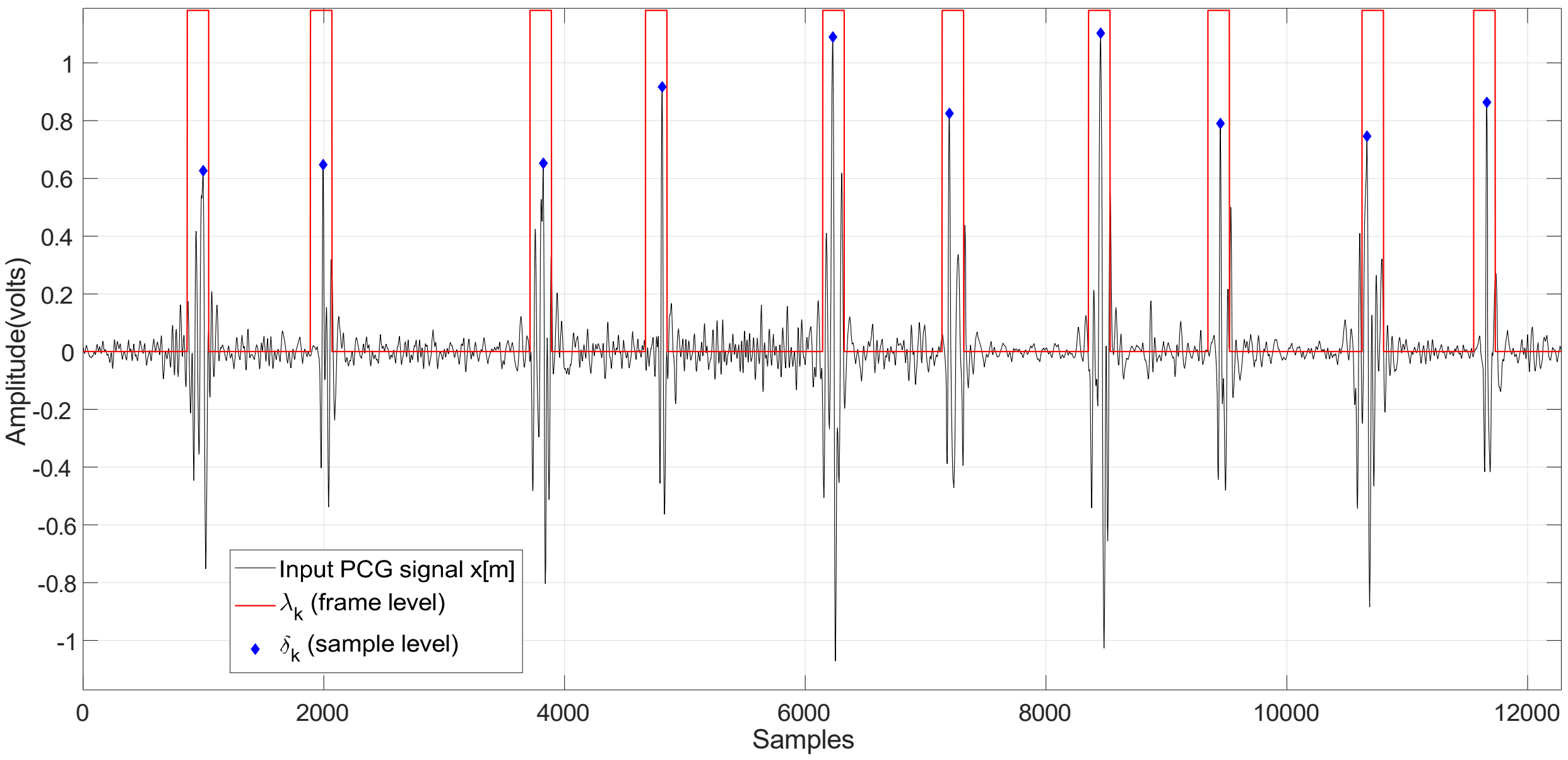

Figure 7: Example of recording "173_1307973611151_C.wav" obtained from database PASCAL [19]. Frame-level and sample-level heartbeat detection. Note that each red rectangle indicates the set of samples $\rho_k$, e.g., the first rectangle is composed of the samples associated to $\rho_1$, the second rectangle is composed of the samples associated to $\rho_2$, etc.

Once the heartbeat detection to a sample-level is performed, the main contribution of this step is the temporal estimation $\beta$, in samples, of the cardiac systole. In this respect, the temporal distance $\xi_d$, in samples, between the previous heartbeat pairs $\delta_k$ is computed as follows,

$$\xi_d = \delta_{k+1} - \delta_k \qquad \forall d, k \in [1, K-1] \tag{4}$$

where $\xi_d$ should contain the distances of the systole and diastole consecutively. Therefore, if the odd distances correspond to the systole, the even distances correspond to the diastole and vice versa (see Figure 8A). Assuming that systole intervals tend to maintain a constant duration over time compared to diastole intervals as previously mentioned, we estimate the cardiac systole duration $\beta$, in samples, by extracting the most repeated distance from $\xi_d$. As can be seen in Figure 8B, the temporal duration of systole $\beta$ is obtained by applying a histogram, composed of $H$ bars, over the vector $\xi_d$. Specifically, $\beta$ is equal to the center value related to the range of the bar with the highest number of repetitions. Experimental results showed that the detection performance remains stable when $H \in [5, 50]$.

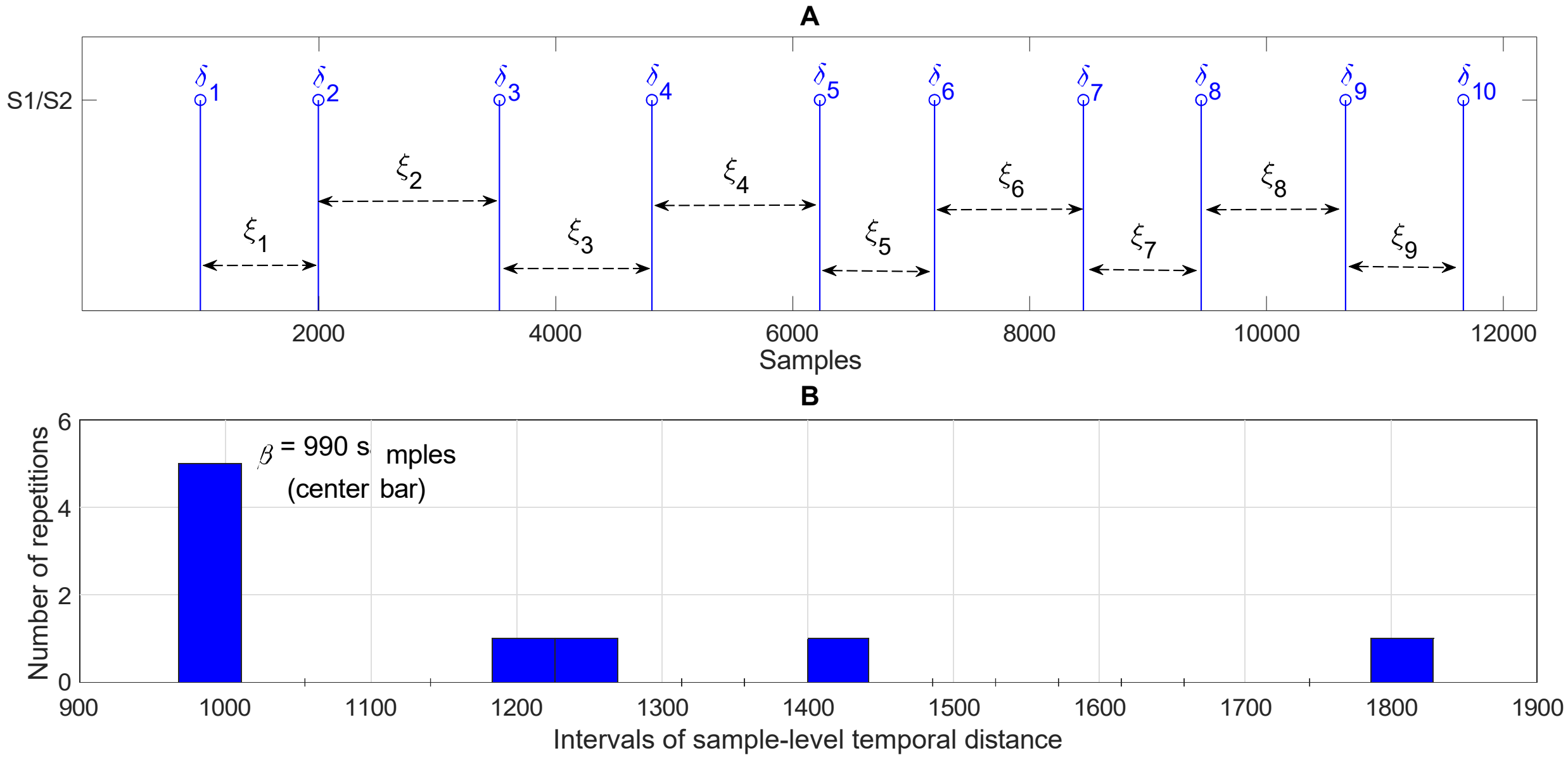

Figure 8: Example of recording "173_1307973611151 C.wav" obtained from database PASCAL [19]. A) Representation of the heartbeats S1/S2 (in a sample-level) by means of $\delta_k$ and the distances $\xi_d$ between the previous heartbeats. B) Histogram of the distance vector $\xi_d$, composed of 20 bars, and the estimation of the systole interval duration $\beta$.

### *2.2.2. Step II: Heart rate estimation and S1/S2 classification*

In order to provide an accurate description of the heartbeats S1/S2, we propose determining the initial heart rate $HR_i$, defined by the two most likely consecutive cardiac cycles assuming the constant temporal duration of systole $\beta$ previously computed, as follows: i) the beat-to-beat distances $\xi_d$ most likely to be systoles are estimated by means of $\phi_s, s = 1, ..., (K-1)/2$, being $\phi_s$ the half of the distances closest to the value $\beta$; ii) the vector $\phi_s$ is sorted in ascending order $\phi_s^{(sorted)}$, that is, a smaller difference between $\xi_d$ and $\beta$ means a higher probability of being a true systole; and iii) a simple search algorithm is applied to locate two systole intervals from the vector $\phi_s^{(sorted)}$ that fulfil the condition that both systole intervals have to be located consecutively within the vector $\xi_d$ and separated by a diastole interval (that is, a distance $\xi_d$ not contained in the vector $\phi^{(sorted)}$). Specifically, the algorithm starts with the most probable systole $\phi^{(sorted)}$ and checks if any of the systoles contained in $\phi_s^{(sorted)}$ satisfy the previous condition in $\xi_d$. In this sense, the algorithm scans the vector $\phi_s^{(sorted)}$ until it finds two systole intervals that fulfil the previous one. Once the two consecutive systole intervals $C1$, $C2$ are located by means of their initial positions $\delta_{S1}^{C1}$, $\delta_{S1}^{C2}$ and final positions $\delta_{S2}^{C1}$, $\delta_{S2}^{C2}$, implicitly, the localization of the associated cardiac cycles is also defined by the states S1, systole, S2 and diastole. Moreover, the initial heart rate $HR_i$ associated to these previous consecutive cardiac cycles can be estimated either as the difference between the initial positions $|\delta_{S1}^{C2} - \delta_{S1}^{C1}|$ or final positions $|\delta_{S2}^{C2} - \delta_{S2}^{C1}|$ because the systole duration is the same for both, as shown in Figure 9.

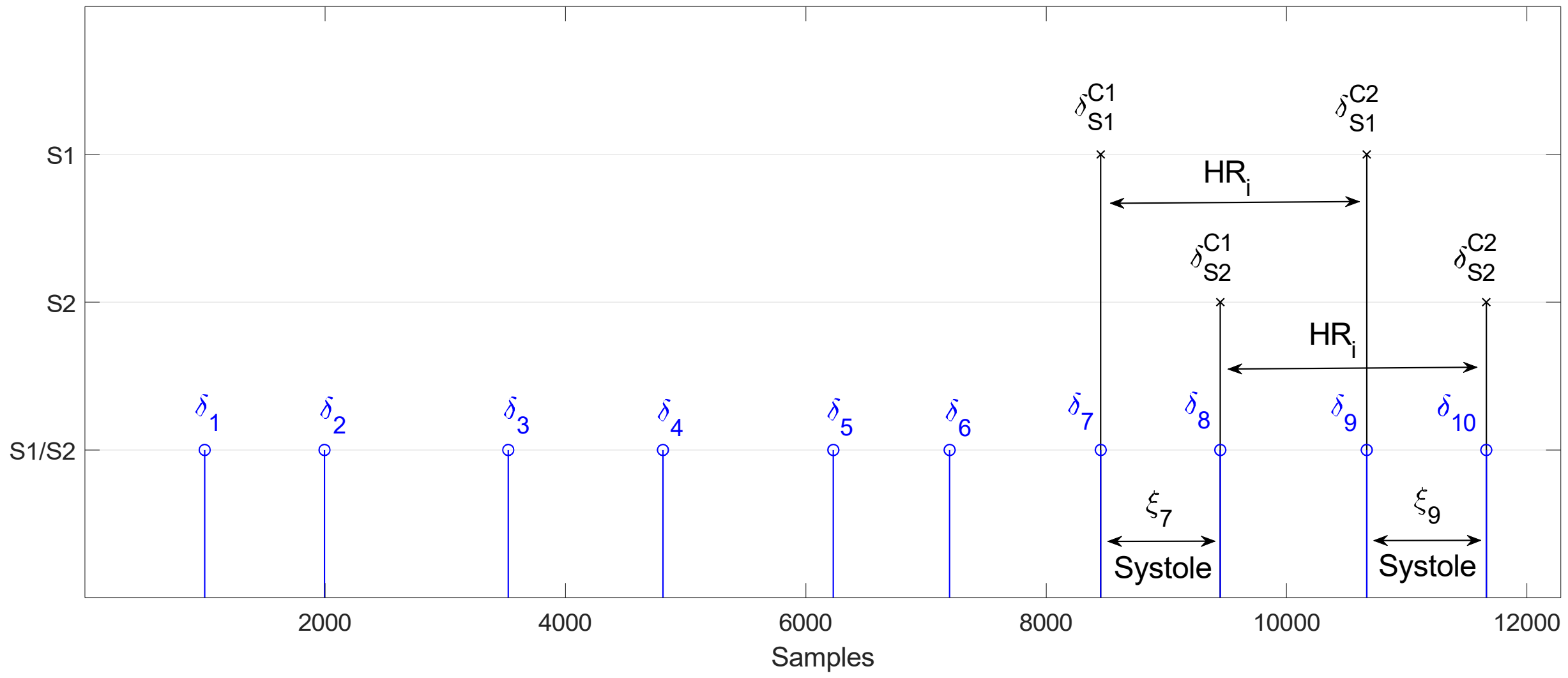

Figure 9: Example of recording "173_1307973611151_C.wav" obtained from database PASCAL [19]. Estimation process of the most likely consecutive systole intervals $C1$, $C2$ based on the constant systole time $\beta$. Note that once both systole intervals have been determined ($\xi_7$ and $\xi_9$), their corresponding heartbeats (S1 and S2) are implicitly labeled. The following information can be extracted at the sample-level: (i) The beginning of the systole intervals: $\delta_{S1}^{C1}$ and $\delta_{S1}^{C2}$; (ii) The final of the systole intervals: $\delta_{S2}^{C1}$ and $\delta_{S2}^{C2}$; (iii) Localization of the four states regarding a cardiac cycle: S1, systole, S2 and diastole and implicitly, the heart rate $HR_i = |\delta_{S1}^{C2} - \delta_{S1}^{C1}| = |\delta_{S2}^{C2} - \delta_{S2}^{C1}|$

Next, a verification-correction algorithm is proposed based on a sliding window that runs through the rest of the previously estimated heartbeats S1/S2 located by the samples $\delta_k$. In this manner, the definitive output of the heartbeat detection is provided by means of the samples $\delta_b$, $b = [1, ..., B]$, being $B$ the number of estimated heartbeats S1/S2 after the validation-correction process has been applied. The main characteristics of the sliding window are the following:

- The window slides from the right to left, starting with the two cardiac cycles associated to the previous systole intervals $C1$, $C2$ that define the initial heart rate $HR_i$, in order to perform a verification-correction of all detected heartbeats $\delta_k$ in the remaining cardiac cycles. Specifically, the window slides to the right starting from the samples $\delta_{S1}^{C2}$ and $\delta_{S2}^{C2}$ belonging to the second estimated cardiac cycle associate to the systole interval $C2$. However, the shift to the left start from the samples $\delta_{S1}^{C1}$ and $\delta_{S2}^{C1}$ belonging to the first estimated cardiac cycle associate to the systole interval $C1$. The process of the window sliding to the left can be observed in Figure 10(A.I), 10(B.I) and 10(C.I).

- The size of the sliding window applied to a given cardiac cycle is variable because the duration of the diastole does not remain constant as previously mentioned. Considering a shift $d$ from the most likely systole intervals $C1$, $C2$, this window size is determined from the value $HR_d$ of the adjacent cardiac cycle. Specifically, the adjacent cardiac cycle to the right is considered when the window slides to the left and vice versa. Following a conservative strategy, the size of the sliding window is delimited using a tolerance margin of error $\eta$, where the lower and upper boundaries of the window range within $[HR_d + \eta, HR_d - \eta]$ from the previous systole interval as shown in Figure 10. Specifically, $\eta$ always has the same value at all window shifts, being used the maximum duration of the heartbeats S1 and S2 in literature. For this reason, we have used $\eta$=160 ms similarly as occurs in [15, 16, 13].

The verification-correction-classification algorithm performs two tasks at each of the sliding window shifts: (I) a verification-correction process applied to heartbeats S1/S2 and then, a heartbeat classification between S1 and S2; and (II) Updating the heart rate $HR_d$ at each shift. Figure 10 reports an example of how the previous sliding window is applied. Subfigures 10A.I, 10B.I, and 10C.I show the verification-correction-classification task, which is explained below. However, Subfigures 10A.II, 10B.II and 10C.II show the updating process regarding to $HR_d$ over successive sliding window shifts. Specifically, a non-fixed value of $HR_d$ along the shifts can be observed due to the temporal variations in diastole.

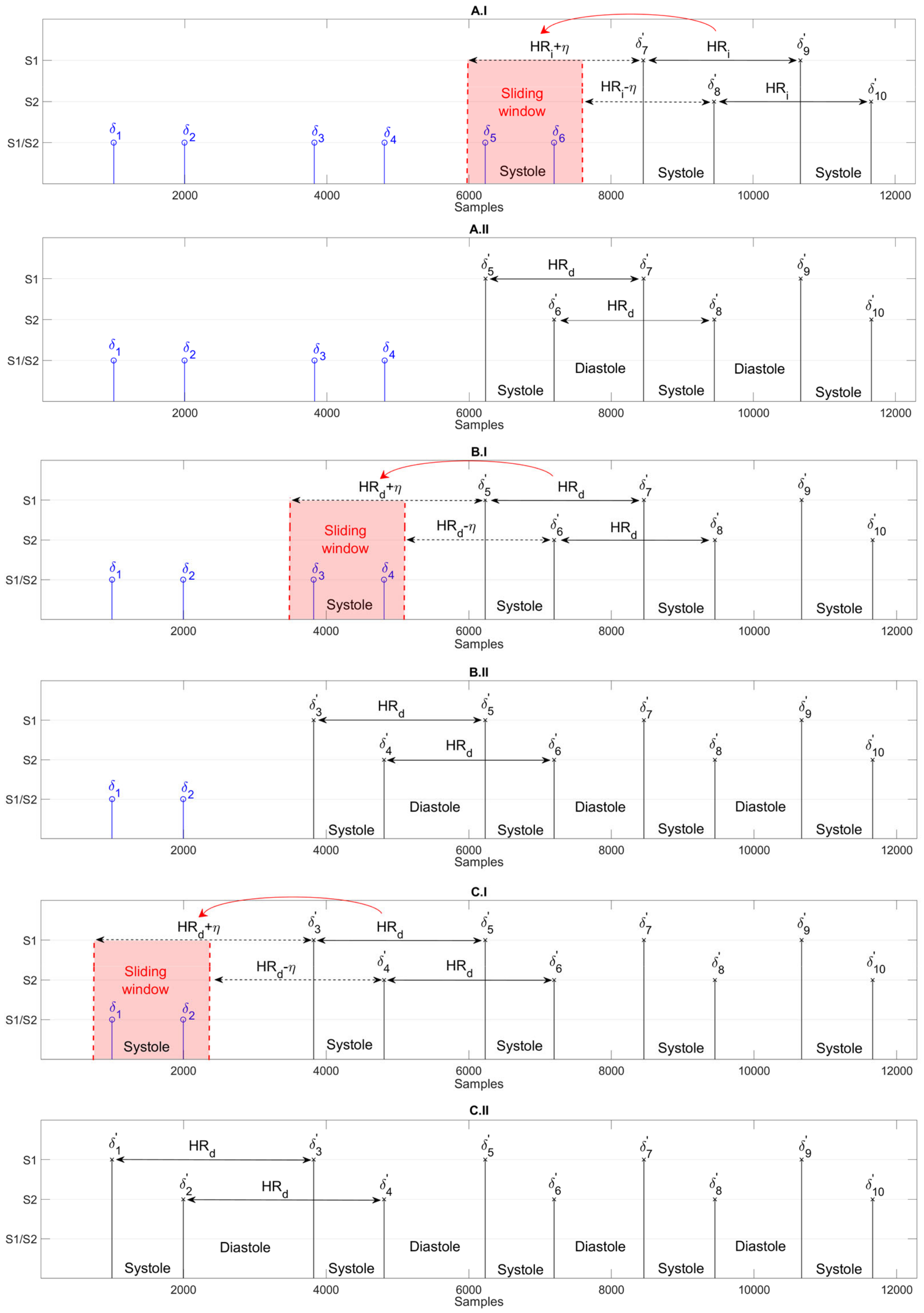

Figure 10: Example of recording ”173_1307973611151_C.wav” obtained from database PASCAL [19]. The method of sliding window considering only the left shift. The subfigures show the window path composed of three shifts: A, B and C. Subfigures *.I show the verification-correction-classification task (from heartbeats S1/S2 to heartbeats S1 or S2). Subfigures *.II show the updating $HR_d$ task. Note that the blue labels $\delta_k$ define the heartbeats S1/S2 before applying the sliding window process but the black labels $\delta'_b$ indicate the heartbeats S1 or S2 after the sliding window procedure has been applied.

In order to apply the verification-correction-classification task on the samples $\delta_k$ at each window shift, we propose to define a set of rules based on some characteristics typically shown by most of the heart sounds to improve the detection and classification of the heartbeats S1/S2: (i) only two heartbeats S1 and S2 can be active in each cardiac cycle; and (ii) the duration of systole has to remain constant over time. Considering the above characteristics, the verification-correction-classification task is defined analysing all scenarios that may occur in each cardiac cycle when it is analysed due to a sliding window shift. Next, we explain the four possible scenarios that can appear depending on the number of heartbeats S1/S2 detected in each cardiac cycle (see Figure 11):

- **Scenario A: no heartbeat has been detected.** A correction stage is applied adding the positions of two new heartbeats S1 and S2 to the vector $\delta_b'$. Specifically, the new heartbeats S1 and S2 are located in the current window at the distance of $HR_d$ from the beginning and final of the adjacent systole interval without applying the margin error $\eta$ in order to maintain the constant systole criterion (see Figure 11A). Besides, the two new heartbeats are labelled S1 and S2 consecutively due to the theoretical structure of a cardiac cycle, composed of the sequence S1, systole, S2 and diastole.
- **Scenario B: only one heartbeat has been detected.** A correction stage is applied adding the position of the missing heartbeat to the vector $\delta_b$. To determine whether the missing heartbeat is S1 or S2, the lower or upper boundary of the window closest to the detected heartbeat $\delta_k$ is identified. For this purpose, the distance of the detected heartbeat $\delta_k$ from the lower boundary $d_L$ and from the upper boundary $d_U$ is obtained. As a consequence, two different situations appear: (i) if the detected heartbeat $\delta_k$ is closer to the lower boundary $d_L < d_U$ (see Figure 11B), the heartbeat is denoted as S1 and the new heartbeat S2, added to the vector $\delta_b'$, is located $+\beta$ samples (systole time) of distance from S1; or (ii) if the detected heartbeat $\delta_k$ is closer to the upper boundary $d_L > d_U$, the heartbeat is denoted as S2 and the new heartbeat S1, added to the vector $\delta_b'$, is located $-\beta$ samples (systole time) of distance from S2;
- **Scenario C: two heartbeats have been detected.** A verification stage is applied to the two detected heartbeats in the vector $\delta_k$ and a result, they are classified as S1 and S2 consecutively (see Figure 11C). Highlight that all shifts shown in Figure 10 belong to Scenario C.
- **Scenario D: more than two heartbeats have been detected.** A correction stage is applied to remove the erroneous heartbeats in the vector $\delta_k$. For this purpose, all possible distances between all possible pairs of detected heartbeats $\delta_k$ in the current cardiac cycle are computed and those whose difference is closest to the systole time $\beta$ is selected, adding the pair of values $\delta_k$ used in the closest difference to the vector $\delta_b'$, being classified as S1 and S2 consecutively (see Figure 11D).

Summarizing, the output of the verification-correction-classification algorithm provides a more reliable heartbeats detection because it solves small errors that may have occurred in Stage I. Moreover, a heartbeats classification discriminating S1 and S2 is achieved, assuming the constant systole duration criterion.

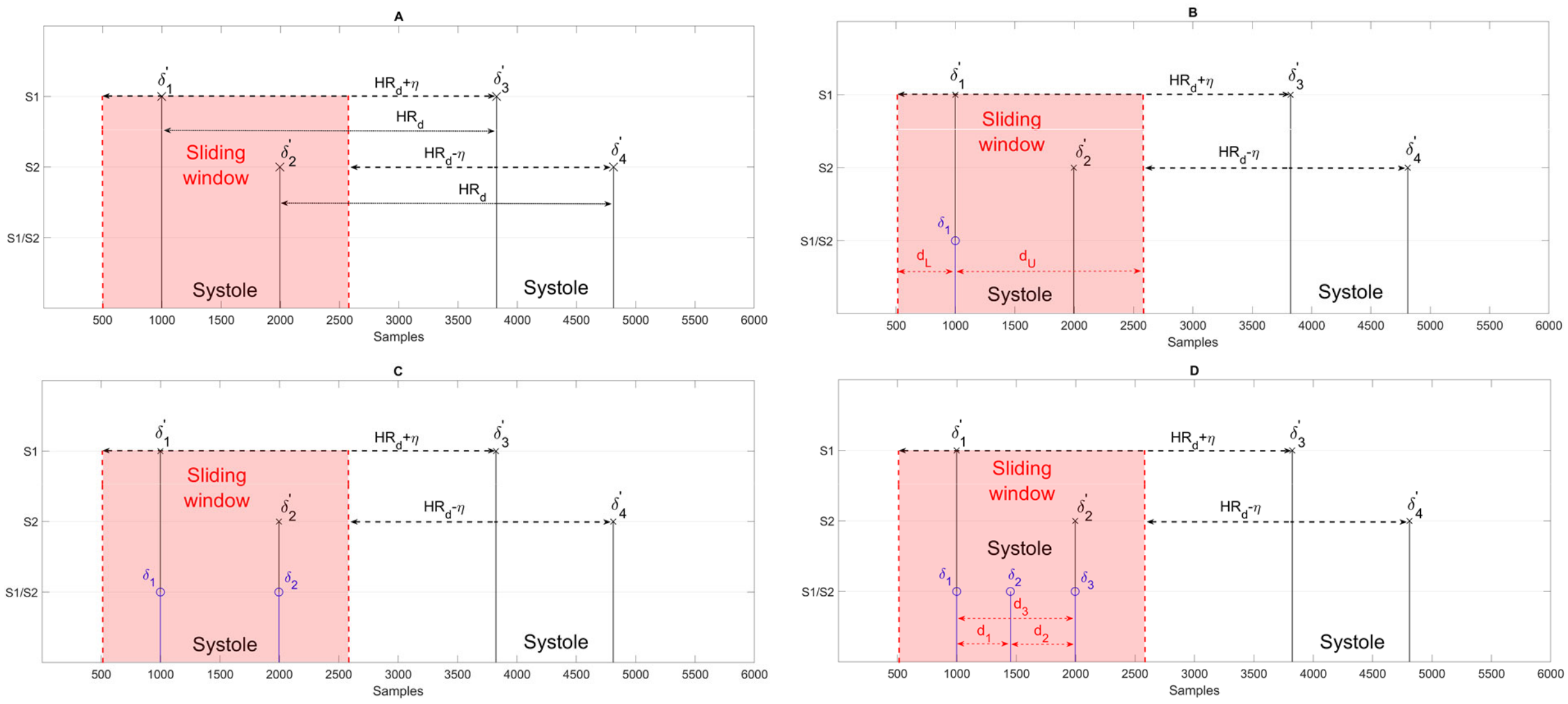

Figure 11: The four scenarios that can be found in the verification-correction-classification task when a sliding window shift is applied. A) Scenario A: no heartbeat has been detected, B) Scenario B: only one heartbeat has been detected, in this example, the heartbeat S1 by means of $\delta_1$. C) Scenario C: two heartbeats have been detected, in this example, the heartbeats S1 and S2 by means of $\delta_1$ and $\delta_2$. D) Scenario D: more than two heartbeats have been detected, being only two of them correct. In this case, the heartbeats by means of $\delta_1$, $\delta_2$, $\delta_3$.

The pseudo-code for Stage II that allows to obtain an improved detection of heartbeats, at the sample level, and heart sound classification between S1 and S2 is shown in Algorithm 2.

---

**Algorithm 2** Fine heartbeat detection

---

**Require**: $x[m]$, $N$, $S$ and $\lambda_k$.

**Step I: Systole duration identification**

1: Localization of the samples S1/S2 by means of $\delta_k$.

2: Compute the temporal distance $\xi_d$ using Equation (4).

3: Compute the cardiac systole duration $\beta$.

**Step II: Heart rate estimation and S1/S2 classification**

4: Compute the two most likely consecutive systole intervals $C1$, $C2$.

5: Compute the initial positions $\delta^{C1}_{S1}$ and $\delta^{C2}_{S1}$ regarding to $C1$, $C2$.

6: Compute the final positions $\delta^{C1}_{S2}$ and $\delta^{C2}_{S2}$ regarding to $C1$, $C2$.

7: Compute the initial heart rate $HR_i$.

8: Compute the lower and upper boundaries $[HR_d + \eta, HR_d - \eta]$ for each window shift.

9: Compute $\delta_b$ applying the verification-correction-classification algorithm on the samples S1/S2 represented by means of $\delta_k$.

**return** $\delta'_b$

---

## 3. Evaluation

### 3.1. Dataset

The detection/classification performance of the proposed method and other recent and relevant state-of-the-art methods has been assessed using the open access dataset PASCAL [19]. Specifically, PASCAL has been widely used in the tasks of segmentation and classification applied to PCG signals [25, 26, 7, 46, 59, 16, 5, 60], since both tasks can be considered as a pre-processing stage for the detection of abnormal heart sounds. The dataset PASCAL is composed of two datasets, A and B, generated by different sources:

- Dataset A: generated from the general public via the iStethoscope Pro iPhone app. This database is composed of 176 heart sound recordings containing both normal and abnormal heart sounds. However, sample-level annotation and S1/S2 classification of heartbeats is only available for 21 of the recordings.

- Dataset B: generated from a clinic trial in several hospitals using the digital stethoscope DigiScope. This database is composed of 656 heart sound recordings, containing both normal and abnormal heart sounds. However, sample-level annotation and S1/S2 classification of heartbeats is only available for 90 of the recordings.

As a consequence, the evaluation performed in this paper has only considered those recordings that have their corresponding heartbeat annotation, resulting in a database $D_T$ composed of 111 recordings, 21 obtained from the dataset A and the remainder 90 recordings from the dataset B.

Additionally, the database $D_T$ has been modified by adding Additive White Gaussian noise (AWGN) with different signal-to-noise ratios (SNR) in order to evaluate the robustness of the heartbeat detection/classification algorithms. In this way, the databases $D_{T_{10}}$ (SNR = 10 dB), $D_{T_5}$ (SNR = 5 dB), $D_{T_0}$ (SNR = 0 dB), $D_{T_{-5}}$ (SNR = -5 dB) and $D_{T_{-10}}$ (SNR = -10 dB) refer to the previous database $D_T$, but using different SNR between each PCG signal $x[m]$ and the noise signal $a[m]$. For example, the database $D_{T_{10}}$ is composed of mixtures in which the power of $x[m]$ is 10 dB greater compared to $a[m]$ and so on for the rest of the databases $D_{T_5}$, $D_{T_0}$, $D_{T_{-5}}$, $D_{T_{-10}}$ taking into account their specific SNR.

To evaluate the detection/classification performance of the proposed method in a realistic sound scenario in which different kinds of clinical ambient noises are active, we have acquired a private database, denoted as $D_N$, through the Soundsnap[1], the world's most widely used sound effects platform. The database $D_N$ is composed of 75 signals recorded in a real clinical environment, with duration between 45 and 307.2 seconds, in which common sounds that appear in a hospital can be heard (e.g., people talking, medical equipment noise, office equipment noise, doors opening and closing, footsteps, etc.). From the clinical noise database $D_N$ and the PCG database $D_T$, the new database $D_C$ has been created by degrading each PCG signal (database $D_T$) with different types of clinical noise that could be found in clinical mediums (database $D_N$). Specifically, the database $D_C$ has been obtained by mixing each recording from the database $D_T$ with a random fragment of one of the noise recordings from the database $D_N$.

Finally, in order to evaluate the robustness of the proposed method when PCG recordings contain abnormal heart sounds representing cardiac abnormalities, the CirCor DigiScope Phonocardiogram Dataset [61, 62] has been used. The CirCor database is composed of 5282 recordings captured from the four main auscultation locations (aortic valve, pulmonary valve, tricuspid valve and mitral valve) of 1568 subjects both with and without cardiac abnormalities. From the CirCor database, we have created the new database $D_O$ by selecting only those PCG recordings in which abnormal heart sounds that represent different common valvular diseases (e.g., aortic stenosis, aortic regurgitation, mitral stenosis, mitral regurgitation, mitral valve prolapse, pulmonary stenosis, tricuspid stenosis, tricuspid regurgitation, septal defects and hypertrophic obstructive cardiomyopathy) are active. Specifically, the new database $D_O$ is composed of 150 recordings with duration between 4.8 and 80.4 seconds.

### *3.2. Metrics*

#### *3.2.1. Heartbeat detection*

Four measures [17, 16, 7], widely used in the field of biomedical signal processing, are used in order to evaluate the proposed method: *Accuracy*, *Precision*, *Recall* and $F1 - score$. All of them are calculated in a sample-by-sample level by comparing the estimated and the ground truth data. An event is considered to be correctly detected when it overlaps in sample-level with the associated event in the ground truth. Note that a maximum time shift of 80 ms in both directions of the event boundaries has been allowed, as the maximum heartbeat duration has been considered to be about 160 ms [15, 16, 13].

[1]https://www.soundsnap.com/tags/clinic

Considering $TP$ as the number of events S1/S2 correctly detected, $TN$ as the number of events non-S1/S2 correctly detected, $FP$ as the number of events non-S1/S2 incorrectly detected and $FN$ as the number of events S1/S2 undetected. Therefore, the four previous metrics are calculated as follows,

$$Accuracy(\%) = \left(\frac{TP+TN}{TP+TN+FP+FN}\right) \cdot 100 \tag{5}$$

$$Precision(\%) = \left(\frac{TP}{TP+FP}\right) \cdot 100 \tag{6}$$

$$Recall(\%) = \left(\frac{TP}{TP+FN}\right) \cdot 100 \tag{7}$$

$$F1-score(\%) = \left(\frac{2TP}{2TP+FP+FN}\right) \cdot 100 \tag{8}$$

Specifically, $Accuracy$ represents the ability to correctly detect the presence or absence of events S1/S2; $Recall$ represents the ability to detect the number of missed events S1/S2 (false negatives); $Precision$ represents the ability to detect the number of events S1/S2 that are inactive in the signal (false positives), and $F1-score$ indicates that both $Precision$ and $Recall$ are of equal importance. Note that the value of $TN$ has not been considered in the computation of the $Accuracy$ to avoid unbalancing the results, as the number of samples where the events S1/S2 are inactive is much higher than the number of samples where the events S1/S2 are active.

Finally, the same metric proposed in [19] has been applied to evaluate the average error associated to the heartbeats localization. Specifically, the metric labelled Total Error (TE) is computed by finding the difference, in milliseconds (ms), of the locations between the estimated one and their corresponding annotated heart sounds,

$$TE_k(ms) = \frac{\frac{\sum_{i=1}^{N_k}(RHS_i - THS_i)}{N_k}}{f_s} \cdot 10^3 \tag{9}$$

where $TE_k$ represents the previous average error associated to the $k^{th}$ PCG recording in the dataset; $RHS_i$ and $THS_i$ is the $i^{th}$ ground truth and detected heartbeat location S1/S2 of the $kth$ PCG recording respectively; $N_k$ is the total number of heartbeats active in the $k^{th}$ PCG recording and $f_s$ is the sampling frequency of the signal. For example, $TE_4 = 20ms$ indicates that the average error for the heartbeats detected in the recording number 4 is 20 ms.

#### 3.2.2. Heartbeat classification

In order to classify an event S1/S2 between S1 or S2, the metrics proposed in [7] have been used: (1) $Accuracy$, which provides the rate of correct classification of both S1 and S2 events; (2) $Sensitivity$, which reports the correct classification of S2 events, that is, the sensitivity is referred as accuracy of class S2; and (3) $Specificity$, which provides the correct classification of S1 events, that is, the specificity is referred as accuracy of class S1.

$$Accuracy(\%) = \left(\frac{TS1+TS2}{TS1+TS2+FS1+FS2}\right) \cdot 100 \tag{10}$$

$$Sensitivity(\%) = \left(\frac{TS2}{TS2+FS2}\right) \cdot 100 \tag{11}$$

$$Specificity(\%) = \left(\frac{TS1}{TS1+FS1}\right) \cdot 100 \tag{12}$$

where $TS1$ is the number of events S1 correctly labelled as S1; $TS2$ is the number of events S2 correctly labelled as S2; $FS1$ is the number of events S1 incorrectly labelled as S2; and $FS2$ is the number of events S2 incorrectly labelled as S1.

### 3.3. Setup

Preliminary results showed the optimal initialisation of parameters in order to obtain the best trade-off between heart sound detection/classification performance and computational cost: sampling rate $f_s$ = 4096 Hz, Hamming window with $N$ = 128 samples length and 25% overlap (temporal resolution of 7.8 ms), a discrete Fourier transform using $2N$ points (frequency resolution of 16 Hz).

### 3.4. State-of-the-art methods for comparison

Two recent state-of-the-art detection/classification heartbeat methods have been used to evaluate the performance of the proposed method: DAS [25] and MUS [26]. Both state-of-the-art methods have been implemented strictly following the modules, stages, block diagram and methodology proposed in their respective manuscripts.

### 3.5. Results

In this section, we assess the potential of the proposed method applied to the heartbeats detection and classification from PCG signals.

#### 3.5.1. Heartbeat detection

The performance of the proposed method and the other state-of-the-art methods in heartbeat detection is evaluated. Focusing on the proposed method, this section also evaluates the heartbeat detection performance provided by the output of Stage I, denoted PM I, and by the output of Stage II, denoted PM II in order to measure the improvement introduced by cascading stages I and II.

Figure 12 shows the detection results evaluating the database $D_T$ when PM_I, PM II and the aforementioned baseline methods DAS and MUS are performed. It reports that both PM_I and PM II provide the best overall detection results, for each metric, compared to the other evaluated methods. Focusing on the average results obtained by PM_I: i) its *Accuracy* improvement is about 11.56% (vs DAS) and 4.12% (vs MUS). It suggests that PM I is more reliable in correctly detecting the presence or absence of cardiac events S1/S2; ii) its *Precision* improvement is about 9.38% (vs DAS) and 1.82% (vs MUS). Thus, results reveal that PM I offers a more robust detection by reducing the number of false positives, that is, events non-S1/S2 incorrectly detected; iii) its *Recall* improvement is about 3.32% (vs DAS) and 3.63% (vs MUS). Therefore, PM I has a more reliable detection by minimising the number of false negatives, that is, events S1/S2 undetected; iv) its $F1 - score$ improvement is about 7.17% (vs DAS) and 2.73% (vs MUS). It can be observed that PM I provides the best balance between false positives and false negatives compared to DAS and MUS. As a consequence, results confirm that the heart temporal structures modelled by the dissimilarity matrix combined with the frame-level spectral divergence (see Section 2.1) is a suitable approach to be applied in heart detection since cardiac sounds show repetitive patterns along time as previously mentioned. Comparing methods DAS and MUS, DAS tends to lose a smaller number of true heartbeats S1/S2 at the expense of increasing the number of false alarms. However, MUS avoids to detect a higher number of false positives related to heartbeats S1/S2 at the expense of losing a higher number of true heartbeats S1/S2.

Figure 12 shows that the best detection performance for all metrics is obtained by PM II. Specifically, PM II achieves a significant improvement of approximately 2.73% (*Accuracy*), 2.03% (*Precision*), 0.81% (*Recall*) and 1.45% ($F1{-}score$) compared to PM I, demonstrating that the verification-correction algorithm used in Stage II of the proposed method (see Section 2.2) improves the heart detection performance by decreasing both the number of false positives and negatives generated in the Stage I of the proposal. A remarkable strength shown by PM_II is that it is the only method that offers a detection performance above 90% for each PCG signal evaluated in the database $D_T$, regardless of the detection metric analysed. Additional results shown in Table 1 have been computed in order to perform a more exhaustive analysis regarding the improvement achieved by PM_II with respect to PM I, measuring the activation rate of the Scenarios used in the verification-correction-classification algorithm by PM_II evaluating the database $D_T$: i) the highest activation rate, 85.85%, is obtained by Scenario C indicating that both heartbeats S1/S2 have been detected by PM I so, PM II only classifies them. In this case, the detection improvement is less than 3% regardless of the metric analysed; ii) the second highest activation rate, 9.05%, corresponds to Scenario

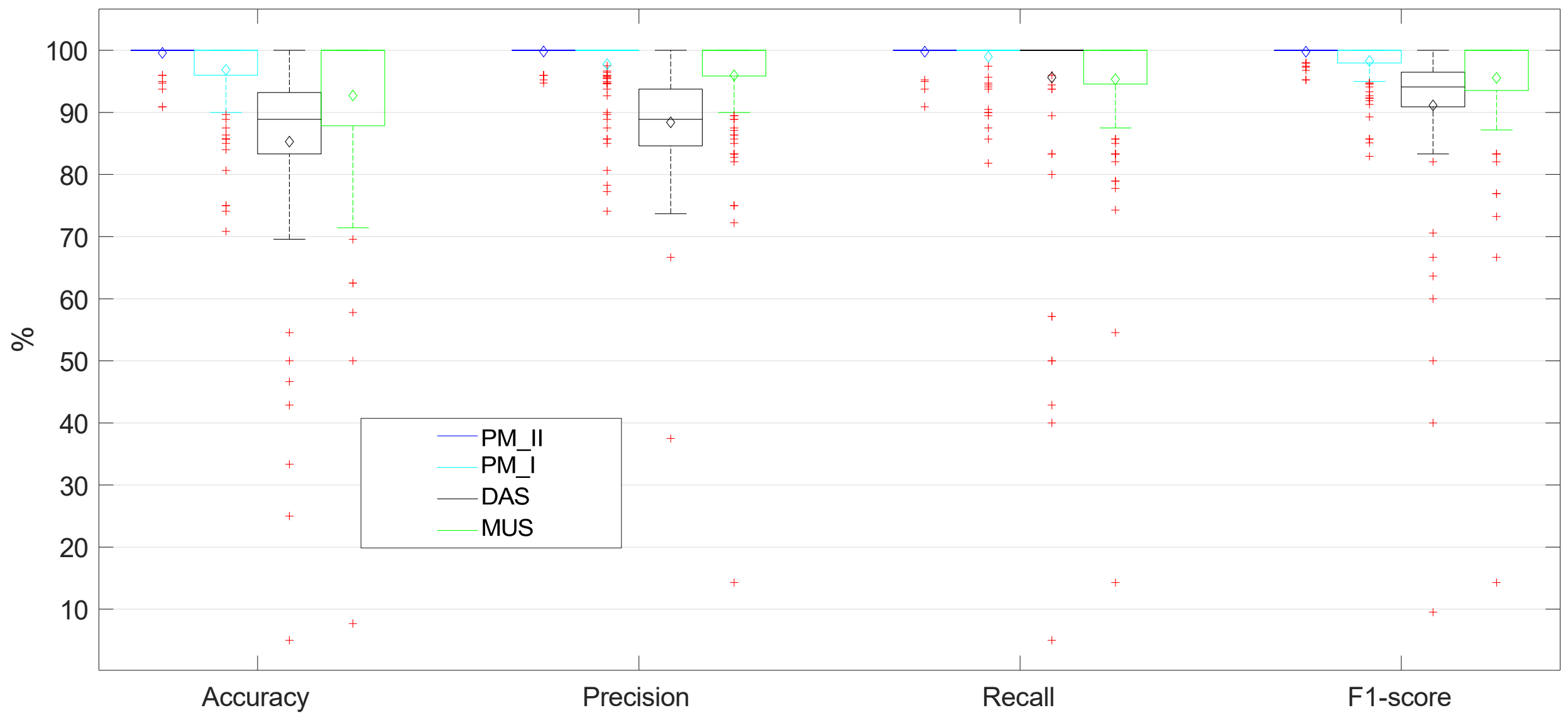

Figure 12: Heartbeats detection results, in term of *Accuracy*, *Precision*, *Recall* and $F1-score$, evaluating the database $D_T$. Each box represents 111 data points, one for each PCG signal of the database $D_T$. The lower and upper lines of each box show the 25th and 75th percentiles. The line in the middle of each box represents the median value. The diamond shape in the center of each box represents the average value. The lines extending above and below each box show the extent of the rest of the samples, excluding outliers. Finally, outliers are defined as points that are over 1.5 times the interquartile range from the sample median, which are depicted as crosses.

D reporting that more than two heartbeats S1/S2 in the cardiac cycle have been previously detected by PM I. Specifically, this scenario helps to decrease the number of false positives eliminating those heartbeats erroneously detected by PM I. In this case, the detection improvement is approximately 2.03% in terms of *Precision*; iii) the third highest activation rate, 5.09%, corresponds to Scenario B in which one heartbeat, without knowing to which type S1 or S2 it belongs, of the cardiac cycle has been previously detected by PM I. This scenario allows to decrease the number of missed heartbeats S1/S2 by adding the temporal location of the heartbeats missed by PM I. The detection improvement is about 0.81% in terms of *Recall*; iv) the lowest activation rate, 0.009%, corresponds to Scenario A, where no heartbeats S1/S2 have been previously detected by PM I. This scenario has only been activated in two cardiac cycles of all PCG signals that compose the database $D_T$. This fact suggests that the probability of finding cardiac cycles without heartbeats detected by PM I is marginal, which confirms the efficiency in heartbeat detection provided by PM I itself.

| Scenario | A | B | C | D |
|---|---|---|---|---|
| Activation rate | 0.009% | 5.091% | **85.850%** | 9.050% |

Table 1: Activation rate of the Scenarios (A, B, C and D) of the verification-correction algorithm of Stage II, belonging to the proposed method PM II evaluating the database $D_T$.

Figure 13 shows the average total error (TE) obtained for each method when the database $D_T$ is evaluated. Results show that the proposed method obtains the lowest TE with a value of less than 10ms per each detected heartbeat, resulting in an improvement of 28ms compared to DAS and 63ms compared to MUS.

Having demonstrated that the proposed method PM II outperforms the heart detection performance compared to PM I, PM II is hereinafter referred to as the proposed method PM.

Figure 14 shows the detection results evaluating the database $D_O$ in which a realistic scenario of PCG signals affected by cardiac abnormalities has been used. The results indicate that the presence of such cardiac abnormalities worsen the detection performance of all evaluated methods as it hinders the identification of heartbeats. Thus, a drop in the average Accuracy value of 9% can be observed for the proposed method, 7% for DAS and 9% for MUS. Specifically, this reduction in detection performance is associated with a remarkable dispersion in Accuracy for all evaluated methods. While the presence of anomalous heart sounds results in a higher detection of inactive S1/S2 cardiac events (false positives) provided by both the proposed method and DAS, in the MUS method there is a higher detection of inactive S1/S2 events together with a higher loss of active S1/S2 events (false negatives). The results seem to suggest that this worsening of detection performance is due to the fact that part of the energy of the abnormal heart sounds is located in the frequency band of heartbeats S1/S2 which makes more accurate detection difficult since some of these abnormal sounds present high spectral similarity to normal sounds S1/S2. Nevertheless, the proposed method still provides a significant improvement over the DAS and MUS methods indicating that the temporal information provided by the dissimilarity matrix together with the proposed verification-correction-classification algorithm is still more accurate and robust than the compared approaches in correctly identifying a more or less repetitive structure that keeps appearing over cardiac cycles in which abnormal heart sounds also appear.

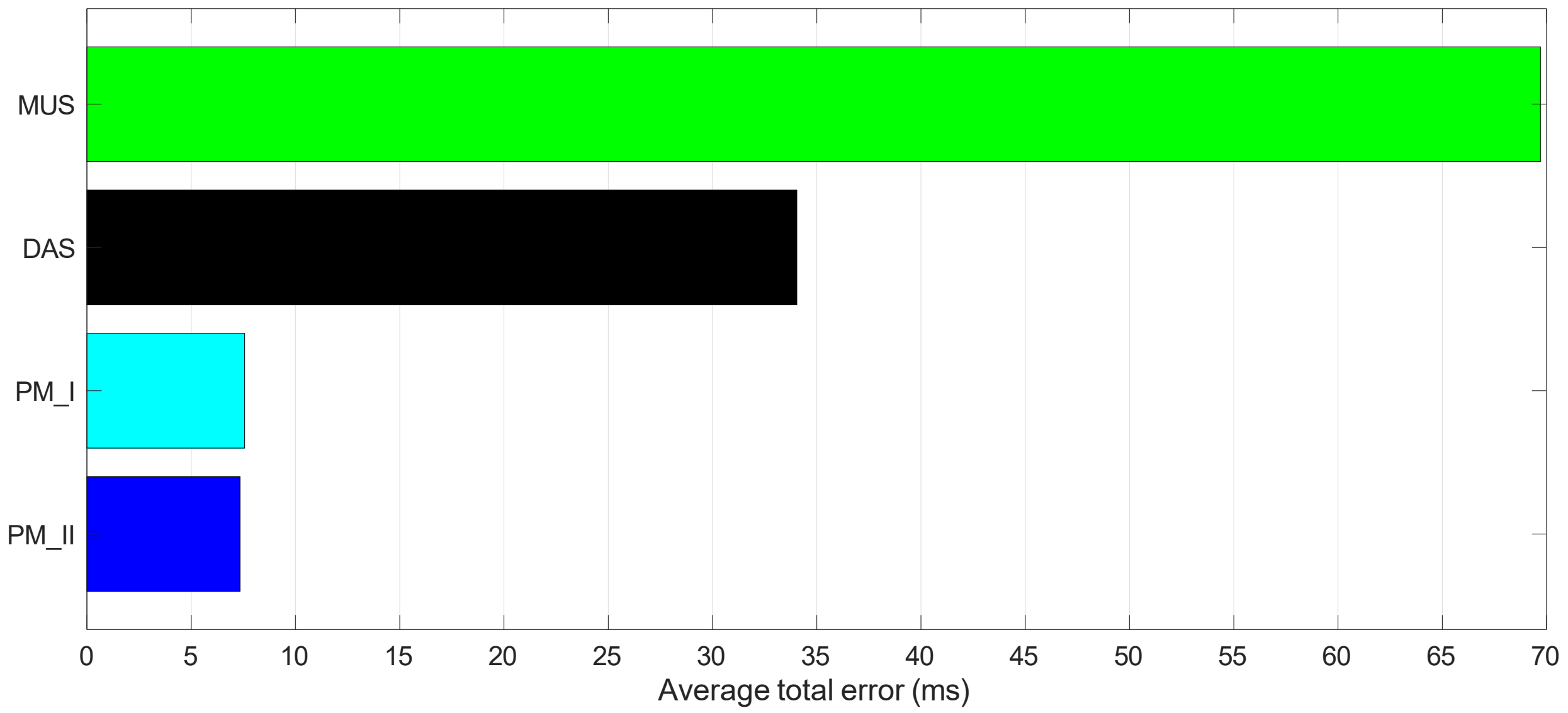

Figure 13: Average total error results, in ms, applied to heartbeat detection evaluating the database $D_T$.

#### *3.5.2. Heartbeat classification*

Figure 15 shows the heartbeat classification results obtained when the database $D_T$ is assessed comparing the performance by the proposed method PM and the aforementioned baseline methods DAS and MUS. Results report that PM provides the best overall classification scores compared to the other evaluated methods considering all classification metrics. Specifically, the proposed method achieves a remarkable improvement of approximately 30.05% (vs DAS) and 23.14% (vs MUS) in terms of average *Accuracy*. Moreover, PM accomplishes a significant improvement of approximately 32.38% (vs DAS) and 25.46% (vs MUS) in terms of average *Specificity* and 29.05% (vs DAS) and 22.09% (vs MUS) in terms of average *Sensitivity*. Although the average performance of the DAS method is worse than that achieved by MUS, it can be observed that both methods minimise both *Specificity* and *Sensitivity* in making errors in classifying both types of heartbeats S1 and S2, being, on average, slightly more inaccurate when classifying S1 events.

In order to determine the strengths and weaknesses of each method, an empirical analysis was also performed, extracting the following conclusions:

- **Focusing on PM**, its classification stage attempts to ensure the correct sequence of the states (S1, systole, S2, diastole) that compose the cardiac cycle. Specifically, the proposed verification-correction-classification algorithm based on sliding window maximises the probability that the structure of the cardiac cycles is maintained throughout the PCG signal, assuming that the duration of systole intervals tends to be constant compared to diastole intervals, avoiding that an event $FP$ or $FN$ occurred in the Stage I does not interfere with the classification results. As a consequence, the classification is highly precise when the identification of the systole duration is correct (see section 2.2.1). However, the proposed method erroneously detects diastole intervals when the duration of these intervals are very similar to systole intervals. As can be seen in Figure 15, the performance obtained by the proposed method is successful since it is always above 82% for all evaluated PCG signals in the database $D_T$, except for only one recording, where an erroneous estimation of the systole duration was performed. Finally, the above results obtained by the proposed method are promising, since these results show a 0.89% probability of erroneously estimating the systole intervals and generating a misclassification.

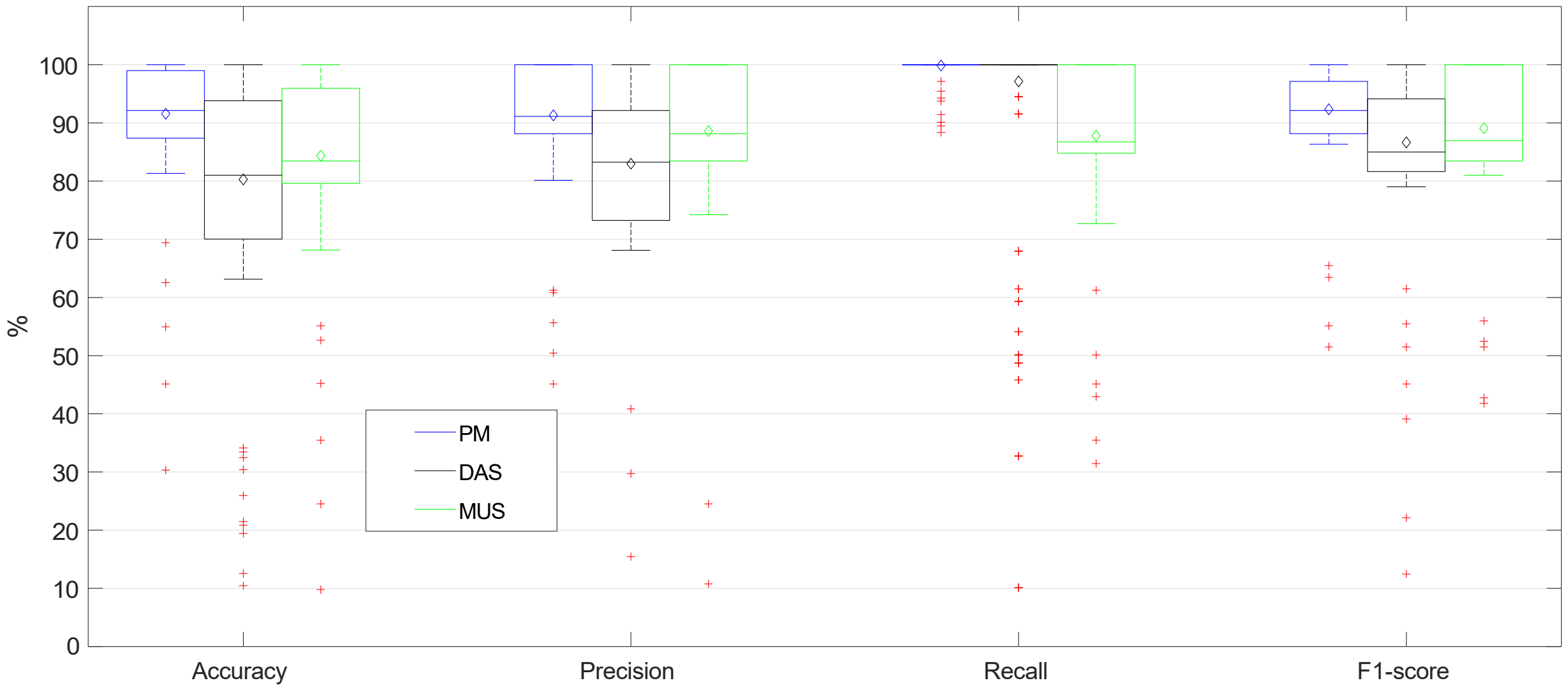

Figure 14: Heartbeats detection results, in terms of *Accuracy*, *Precision*, *Recall* and $F1 - score$, evaluating the database $D_O$. Each box represents 150 data points, one for each PCG signal of the database $D_O$. The lower and upper lines of each box show the 25th and 75th percentiles. The line in the middle of each box represents the median value. The diamond shape in the center of each box represents the average value. The lines extending above and below each box show the extent of the rest of the samples, excluding outliers. Finally, outliers are defined as points that are over 1.5 times the interquartile range from the sample median, which are depicted as crosses.

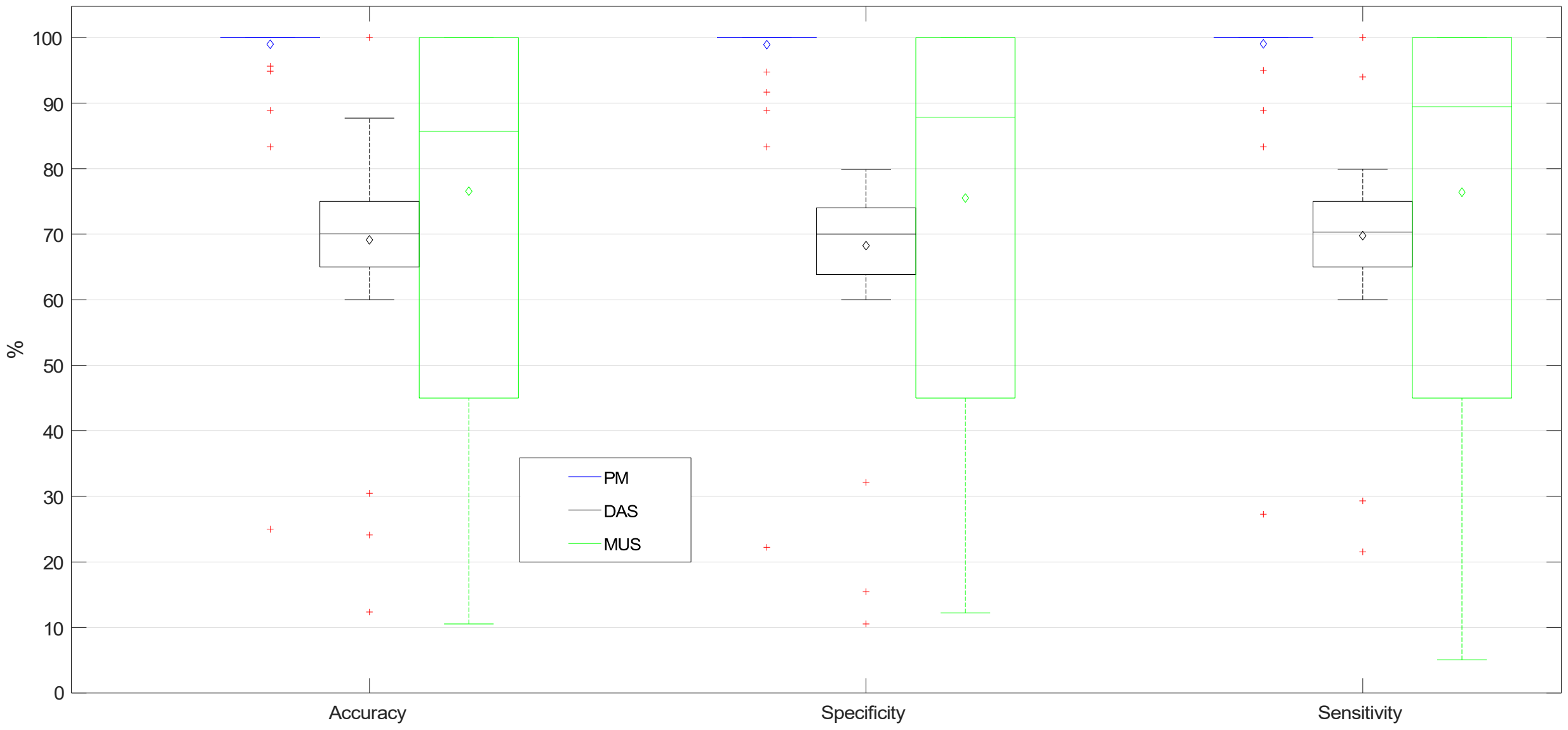

Figure 15: Heart classification results, in terms of *Accuracy*, *Specificity* and *Sensitivity*, evaluating the database $D_T$. Each box represents 111 data points, one for each PCG signal of the database $D_T$. The lower and upper lines of each box show the 25th and 75th percentiles. The line in the middle of each box represents the median value. The diamond shape in the center of each box represents the average value. The lines extending above and below each box show the extent of the rest of the samples, excluding outliers. Finally, outliers are defined as points that are over 1.5 times the interquartile range from the sample median, which are depicted as crosses.

- **Focusing on MUS**, the high variance, approximately 90%, shown by the results is due to its weak classification criterion. Specifically, its classification stage divides the detected heartbeats into two groups, odd and even, and labels the group with higher energy as S1 and the other as S2 [26]. In this manner, the classification results are efficient when the detection stage is correct and no $FP$ or $FN$ appears (in other words, if all heartbeats have been detected, then they are correctly classified). However, the performance is drastically reduced when the heartbeat detection estimates one or more than one $FP$ or $FN$. For example, supposing that a given PCG signal is composed of four heartbeats S1, S2, S1 and S2, and the algorithm generates a $FP$ detecting S1, $FP$, S2, S1 and S2, neither the odd elements correspond to S1 heartbeats, nor the even elements to S2 heartbeats.

- **Focusing on DAS**, the main weakness is due to its classification stage does not implement any post-processing stage to ensure the common structure of the events belonging to a cardiac cycle since it does not follow the alternation between the occurrence of two types of heartbeats S1 and S2. As a consequence, DAS allows several consecutive heartbeats to be labelled as the same type S1 or S2 [25].

Figure 16 shows the results of the classification evaluating the database $D_O$ in which a realistic scenario of PCG signals affected by cardiac abnormalities has been used. Again, the results indicate that the presence of such cardiac abnormalities reduces the classification performance of all the methods evaluated, however, the observed reduction in the average value of Accuracy, as well as the increase in the dispersion associated with Accuracy, is greater in the classification task compared to the detection task for all methods compared. However, and as occurred when evaluating PCG signals without abnormal heart sounds, the lowest dispersion is obtained by the proposed method, followed by DAS and finally MUS, which presents the highest dispersion. Specifically, this worsening of classification performance is due, with approximately equal importance, to the errors made in the classification of the S1 and S2 events obtained by the proposed method and DAS, while the errors made by MUS mainly affect the S1 events. It should be noted that the remarkable dispersion behavior obtained by the proposed method confirms that the set of rules on which the proposed verification-correction-classification algorithm is based is capable of correctly segmenting and classifying most PCG signals, including those with presence of abnormal heart sounds, since the algorithm eliminates most of the spurious S1/S2 events and recovers the missed S1/S2 events in previous stages, guaranteeing the correct sequence of the states S1, systole, S2, diastole that compose the structure of a cardiac cycle.

### *3.5.3. Heartbeat detection/classification under simulated noisy conditions*

Table 2 shows *Accuracy*, *Precision*, *Recall* and $F1 - score$ results in order to evaluate the heartbeat detection robustness of the proposed method PM and the baseline methods DAS and MUS using five different SNR datasets: $D_{T_{10}}$, $D_{T_5}$, $D_{T_0}$, $D_{T_{-5}}$ and $D_{T_{-10}}$. Experimental results indicate that PM provides the best overall detection performance compared to the other evaluated baseline methods considering all SNR scenarios assessed. Focusing on the worst evaluated acoustic scenario $D_{T_{-10}}$ in which the AWGN noise is louder than the heart sounds, it can be observed an improvement about 14.20% (vs DAS) and 11.89% (vs MUS) in *Accuracy*, 4.93% (vs DAS) and 6.79% (vs MUS) in *Precision*, 9.62% (vs DAS) and 6.29% (vs MUS) in *Recall* and finally, 7.07% (vs DAS) and 5.56% (vs MUS) in $F1 - score$. Results confirm that PM is the most effective and reliable method for detecting heartbeats under AWGN noisy scenarios, providing a performance of over 94% for all metrics and SNR analysed. This robustness shown by PM can be considered a promising ability to detect the presence of weak heartbeats that can be acoustically masked in noisy environments.

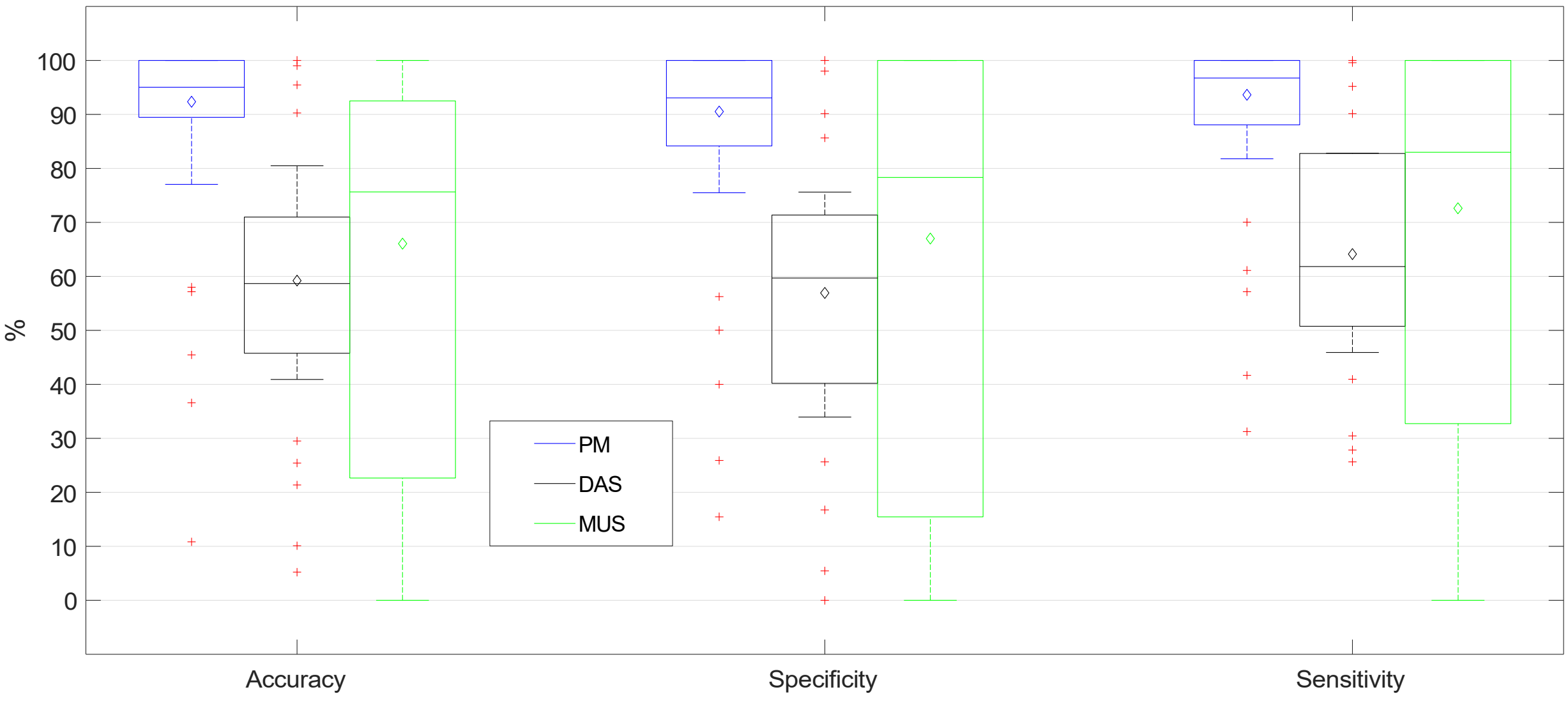

Figure 16: Heart classification results, in terms of *Accuracy*, *Specificity* and *Sensitivity*, evaluating the database $D_O$. Each box represents 150 data points, one for each PCG signal of the database $D_O$. The lower and upper lines of each box show the 25th and 75th percentiles. The line in the middle of each box represents the median value. The diamond shape in the center of each box represents the average value. The lines extending above and below each box show the extent of the rest of the samples, excluding outliers. Finally, outliers are defined as points that are over 1.5 times the interquartile range from the sample median, which are depicted as crosses.

Comparing the detection results from the datasets $D_{T_{10}}$ and $D_{T_{-10}}$ that simulate the most harmful and beneficial noisy scenario, they show a performance reduction of about 4.51% (PM), 5.75% (DAS) and 9.91% (MUS) in *Accuracy*, 4.67% (PM), 8.8% (DAS) and 8.6% (MUS) in *Precision*, 1.66% (PM), 7.93% (DAS) and 3.61% (MUS) in *Recall* and 4.37% (PM), 4.63% (DAS) and 5.93% (MUS) in $F1 - score$. Note that PM suffers the smallest drop in performance compared to the other state-of-the-art methods, at best with a reduction by a factor of 4. As a result, it can be stated that the proposed method using temporal information from the dissimilarity matrix applied to the localization/classification of the heartbeats is an efficient and reliable tool, even when the PCG signal can be masked with AWGN noise that overlaps the same spectral bands at the same time of the heart sounds.

Table 3 shows *Accuracy*, *Specificity* and *Sensitivity* classification results, evaluating the previous five SNR datasets to compare the robustness of the S1/S2 heartbeat classification task of the proposed method and the state-of-the-art methods. The results report that PM obtains the best classification results regardless of the metric and for all the SNR scenarios evaluated. Specifically, PM is the only method that achieves that its worst performance is above 90% in all the evaluated datasets while DAS and MUS obtain results below 80% in all the SNR scenarios, even with values below 72% when SNR<0dB and implicitly, a greater time-frequency overlapping between noise and heart sounds occurs. Due to both the above results and the fact that PM reduces its performance by a smaller percentage than DAS and MUS as the SNR of the simulated scenario decreases, PM confirms its best heart classification performance in noisy environments. Thus, PM reduces its average *Accuracy* performance by approximately 7% while DAS and MUS reduce it by 9% and 11% when comparing the most harmful (SNR=-10dB) and beneficial (SNR=10dB) acoustic environments evaluated.

Figure 17 and Figure 18 show the detection and classification results evaluating the database $D_C$ in which a realistic scenario of PCG signals affected by different types of clinical environmental noise has been created. As occurs when evaluating PCG signals in which abnormal heart sounds appear, the presence of different types of ambient noises that can be found in clinical settings worsen the detection and classification performance of all the methods evaluated. This reduction in detection and classification affects the proposed method to a lesser extent, which continues to significantly exceed the performance shown by DAS and MUS. However, in all the methods evaluated, the drop in performance in the classification task is more notable. Focusing on detection, the presence of clinical ambient noise causes both the proposed method and DAS the same performance when no ambient noises are active, while MUS increases the creation of spurious heart events and the loss of heart events S1 and S2, which could be considered more critical for future analysis of heart sound abnormalities. It is noteworthy that while DAS and MUS significantly increase the dispersion of the Accuracy results in detection and classification, higher in classification (8% with respect to DAS and 10% with respect to MUS) compared to detection, the proposed method maintains the dispersion values approximately constant with respect to the scenario without interference from ambient noise. It should be noted that the high dispersion obtained by MUS is mainly due to the fact that it bases its classification on a criterion based on the energy levels of heart sound events, a poor criterion when evaluating realistic environments with ambient noise interference since the acoustic mixture of clinical ambient noise could often cause the presence of heart events S2 with greater energy than events S1. Although DAS ranks worst in terms of classification at the expense of showing lower dispersion compared to MUS, its classification performance could be improved by adding a control algorithm that would allow discerning the occurrence of a new heart event S1 or S2 as long as the cardiac temporal repetitive structure is maintained.

| Algorithm | *Accuracy* (%) | *Precision* (%) | *Recall* (%) | $F1-score$ (%) |
|---|---|---|---|---|
| | | Database $D_{T_{10}}$ (SNR = 10dB) | | |
| PM | **99.38** | **99.63** | **99.69** | **99.66** |
| DAS | 86.41 | 89.00 | 96.33 | 91.83 |
| MUS | 92.88 | 96.15 | 95.34 | 95.65 |
| | | Database $D_{T_5}$ (SNR = 5dB) | | |
| PM | **99.37** | **99.62** | **99.69** | **99.65** |
| DAS | 85.18 | 88.75 | 95.12 | 91.03 |
| MUS | 92.79 | 95.95 | 95.30 | 95.54 |
| | | Database $D_{T_0}$ (SNR = 0dB) | | |
| PM | **99.11** | **99.47** | **99.57** | **99.51** |
| DAS | 84.63 | 88.05 | 93.39 | 90.51 |
| MUS | 92.44 | 95.53 | 95.08 | 95.40 |
| | | Database $D_{T_{-5}}$ (SNR = -5dB) | | |
| PM | **97.16** | **97.61** | **99.12** | **98.09** |
| DAS | 82.27 | 87.56 | 91.26 | 88.97 |
| MUS | 87.54 | 92.92 | 92.41 | 92.50 |
| | | Database $D_{T_{-10}}$ (SNR = -10dB) | | |
| PM | **94.87** | **94.96** | **98.02** | **95.28** |
| DAS | 80.66 | 85.03 | 88.40 | 87.20 |
| MUS | 82.97 | 88.17 | 91.73 | 89.71 |

Table 2: Detection results, in terms of *Accuracy*, *Precision*, *Recall* and $F1-score$, evaluating the databases $D_{T_{10}}$, $D_{T_5}$, $D_{T_0}$, $D_{T_{-5}}$ and $D_{T_{-10}}$.

## 4. Discussion

In this study, we addressed the tasks of detection and classification in single-channel PCG recordings using an unsupervised approach based on time-frequency behaviors shown by most of heartbeats. From a medical point of view, the analysis of PCG signals allows a low-cost, non-invasive and reliable diagnosis to be applied in poor-resources areas. From an engineering point of view, detection and classification using signal processing in PCG signals can help the clinician to better interpret the meaning of heart sounds heard through medical devices, as well as be used as a pre-processing tool to develop algorithms to identify the presence of cardiac diseases by learning the target spectral content located in specific temporal intervals localized by the previous tasks.

| Algorithm | *Accuracy* (%) | *Specificity* (%) | *Sensitivity* (%) |
|---|---|---|---|
| Database $D_{T_{10}}$ (SNR = 10dB) | | | |
| PM | **97.18** | **97.12** | **97.24** |
| DAS | 70.29 | 69.01 | 70.76 |
| MUS | 77.59 | 78.24 | 79.66 |
| Database $D_{T_5}$ (SNR = 5dB) | | | |
| PM | **95.53** | **95.47** | **95.59** |
| DAS | 68.49 | 67.79 | 68.38 |
| MUS | 75.00 | 76.99 | 77.00 |
| Database $D_{T_0}$ (SNR = 0dB) | | | |
| PM | **92.68** | **92.62** | **92.74** |
| DAS | 67.47 | 66.88 | 67.46 |
| MUS | 72.33 | 73.54 | 74.55 |
| Database $D_{T_{-5}}$ (SNR = -5dB) | | | |
| PM | **91.49** | **91.89** | **91.39** |
| DAS | 64.98 | 64.00 | 65.07 |
| MUS | 70.00 | 71.05 | 71.99 |
| Database $D_{T_{-10}}$ (SNR = -10dB) | | | |
| PM | **90.79** | **91.10** | **90.51** |
| DAS | 61.41 | 60.90 | 61.00 |
| MUS | 66.78 | 67.49 | 68.64 |

Table 3: Classification results, in terms of *Accuracy*, *Specificity* and *Sensitivity*, evaluating the databases $D_{T_{10}}$, $D_{T_5}$, $D_{T_0}$, $D_{T_{-5}}$ and $D_{T_{-10}}$.

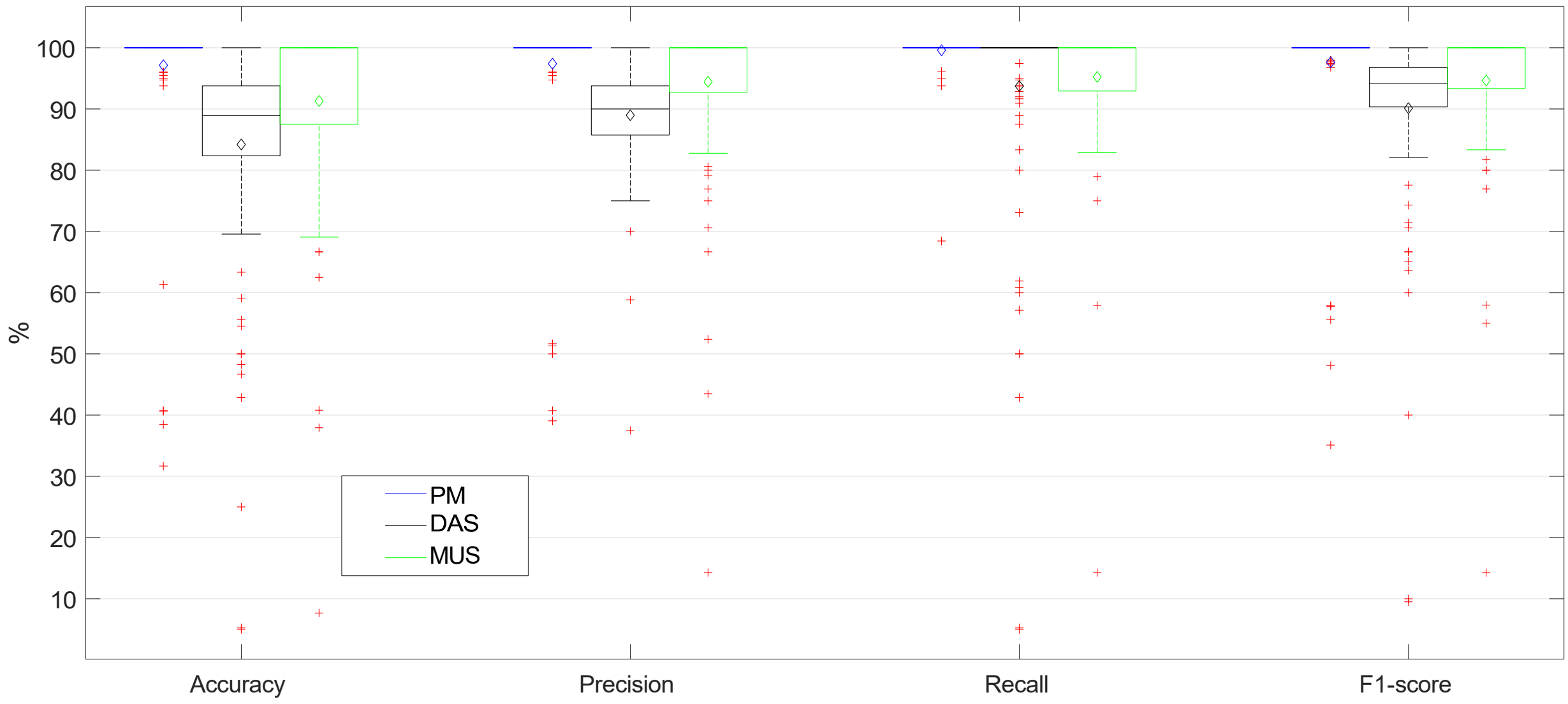

Figure 17: Heartbeats detection results, in terms of *Accuracy*, *Precision*, *Recall* and $F1-score$, evaluating the database $D_C$. Each box represents 111 data points, one for each PCG signal of the database $D_C$. The lower and upper lines of each box show the 25th and 75th percentiles. The line in the middle of each box represents the median value. The diamond shape in the center of each box represents the average value. The lines extending above and below each box show the extent of the rest of the samples, excluding outliers. Finally, outliers are defined as points that are over 1.5 times the interquartile range from the sample median, which are depicted as crosses.

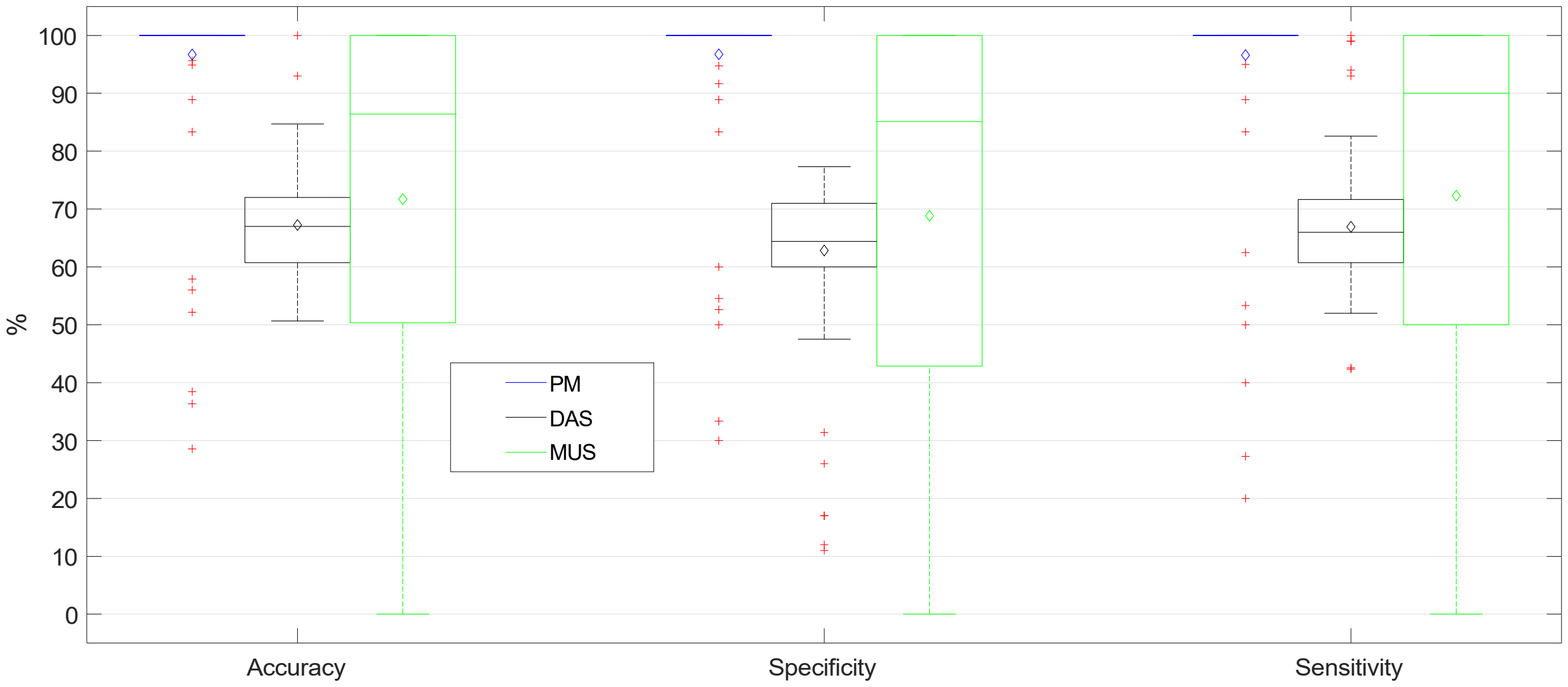

Figure 18: Heart classification results, in terms of *Accuracy*, *Specificity* and *Sensitivity*, evaluating the database $D_C$. Each box represents 111 data points, one for each PCG signal of the database $D_C$. The lower and upper lines of each box show the 25th and 75th percentiles. The line in the middle of each box represents the median value. The diamond shape in the center of each box represents the average value. The lines extending above and below each box show the extent of the rest of the samples, excluding outliers. Finally, outliers are defined as points that are over 1.5 times the interquartile range from the sample median, which are depicted as crosses.

The detection and classification tasks described in this study have been specially developed to analyze the temporal structure exhibited by most heart sounds. Of the two cascaded stages of which the proposed system is composed, the first stage attempts to detect S1 and S2 heart sounds by assuming a temporally repetitive cardiac spectral structure where similar spectral patterns are observed between S1 and S2 but different from the patterns that can be found in systole and diastole and, in addition, the duration of both S1 and S2 is shorter than systole and diastole. For that, we propose to model such temporal structure using the dissimilarity matrix that exploits the aforementioned behaviors in most heart sounds. To improve the temporal localization of the heartbeats obtained in the first stage and subsequently classify them as S1 or S2, the second stage presents a verification-correction-classification algorithm assuming that only two heart sounds S1 and S2 can be active in each cardiac cycle and the duration of systole can be considered constant over time.

An exhaustive assessment has been carried out, analyzing PCG signals with and without the interference caused by abnormal heart sounds and different kinds of noise such as clinical ambient noise, in order to evaluate both the detection and classification performance of the proposed method. Results indicated a significant detection and classification improvement of the proposed method compared to the state-of-the-art methods evaluated, showing that the use of the dissimilarity matrix to extract temporal information from PCG signals is a promising tool to be applied in heart segmentation and further, to be applied as a pre-processing stage to heart sound signal processing. While a notable strength shown by the proposed method is its ability to update the heart rate by analysing a new cardiac cycle within the time structure corresponding to the entire PCG signal, in addition, it is the only method evaluated that provides a detection performance above 90% for each PCG signal in the database $D_T$. However, a weakness of the proposed method is that it often erroneously detects a diastole interval when the duration of this interval is very similar to the duration of a systole interval. The verification-correction-classification algorithm significantly minimizes the number of false positives and false negatives initially detected because all the possible heart sound scenarios, that can appear depending on the number of heartbeats detected in the first stage, can be corrected.

Considering both the presence of abnormal heart sounds and the different types of ambient noises that can be found in clinical environments, it is observed as a general behavior that all methods evaluated worsen their detection and classification results. This performance reduction affects to a lesser extent the proposed method, which still significantly outperforms DAS and MUS. Focusing on the acoustic interference caused by the presence of clinical ambient noise, it is noteworthy that while DAS and MUS significantly increase their dispersion of the Accuracy results in detection and classification, the proposed method maintains its dispersion values approximately constant with respect to the scenario without interference from ambient noise. As a result, our approach can be considered reliable even when the PCG signal is masked with different types of noise which appear superimposed at the same time as the heart sounds as well as in the same spectral band in which the heartbeats concentrate most of their energy. Focusing on the acoustic interferences caused by the presence of abnormal heart sounds, it can be observed that the set of rules on which the proposed verification-correction-classification algorithm is based correctly segments and classifies most of the PCG signals including those with the presence of abnormal heart sounds, as the algorithm allows to eliminate a large number of spurious S1/S2 events as well as to recover most of the missing S1/S2 events. It can be considered another strength shown by our proposal since these facts ensure a correct, time-ordered sequence of the S1, systole, S2, diastole states that make up the structure of a cardiac cycle.

## 5. Conclusions and Future work

In this work, we propose an unsupervised approach to detect and classify heartbeats from PCG signals. In order to simulate the temporal-spectral behaviour of heart sounds, the proposed method assumes the following hypotheses: i) heart sounds can be considered as spectral structures that repeat over time; ii) S1 and S2 heartbeats show similar spectral patterns to each other, but sufficiently different from those that can be contained in the systole and diastole; iii) the duration of S1 and S2 heartbeats is always shorter than systole and diastole; and iv) systole intervals tend to maintain a constant duration over time compared to diastole intervals. In the detection stage, our main contribution proposes a novel version of the standard similarity matrix, named dissimilarity matrix, combined with the frame-level spectral divergence in order to estimate heartbeats assuming both the repetitiveness typically shown by the heart sounds and temporal relationships between events S1/S2 and non-S1/S2 (systole and diastole). In the classification stage, our main contribution develops a verification-correction-classification method, based on a sliding window, that allows maintaining the structure of the cardiac cycle throughout the PCG signal by means of S1, systole, S2 and diastole. The most relevant conclusions obtained in the experimental results are detailed below:

- The proposed method obtains the best performance in heartbeat detection, providing average results above 99% and 92% for PCG signals without and with cardiac abnormalities, regardless of the metrics analysed. In this sense, the results suggest that the dissimilarity matrix is an effective tool for estimating heartbeats from the PCG signal.

- The proposed method obtains the best classification performance for S1 and S2 heartbeats, offering average results higher than 97% and 92% for PCG signals without and with cardiac anomalies, regardless of the metric analysed. In this sense, the results indicate that the algorithm based on sliding window tracking to maintain the structure of the cardiac cycle is reliable to be applied in the segmentation of the heartbeats of each cycle.

- Similar to what occurs in the real world, detecting and classifying heartbeats is much more complex in acoustic scenarios in which heart sounds are barely audible due to the time-frequency overlap of some kind of acoustic interference, such as AWGN noise. In this respect, highlight that the detection/classification results of the proposed method show the smallest drop compared to the other state-of-the-art methods. Specifically, comparing the evaluation of the databases $D_{T_{10}}$ and $D_{T_{-10}}$, the detection results provided by the proposed method drop below 4.5% (DAS drops below 5.8% and MUS 9.9%) and the classification results drop below 6.7% (DAS drops below 8.9% and MUS 10.8%), in terms of *Accuracy*. It suggest that the proposed method is more robust than baseline methods in the assessment of noisy scenarios, specifically, SNR<0dB. Moreover, considering a realistic scenario in which different kinds of clinical ambient noises are active, the proposed method obtains the best heartbeat detection/classification performance, providing average results above 96% regardless of the metric employed. These results indicate that the proposed method is the most robust method among those evaluated when real clinical scenarios are assessed.

- Finally, the proposed method estimates the lowest temporal error compared to the other state-of-the-art methods. Specifically, the proposed method produces an error of less than 10 ms per each detected heartbeat, while the other methods exceed 34 ms.

Future work will focus in two directions: (i) normal/abnormal heart classification using heart segmentation provided by the proposed method that will be used as a pre-processing stage, and (ii) development of novel time-frequency representations that improve the extraction of features to be applied to neural network architectures in order to classify major CVDs, such as mitral stenosis and mitral regurgitation.

## Funding

This work was supported by the Programa Operativo FEDER Andalucia 2014-2020 under project with reference 1257914, the Ministry of Economy, Knowledge and University, Junta de Andalucia under Project P18-RT-1994 and the Spanish Ministry of Science and Innovation under Project PID2020-119082RB-C21.

## Conflict of interest

The authors have no conflict of interest to disclose. The authors are responsible for the content and writing of this article alone.

## Acronyms

| | |
|---|---|
| AWGN | Additive White Gaussian Noise |
| BPM | Beats Per Minute |
| CNN | Convolutional Neural Network |
| CVD | Cardiovascular Disease |
| ECG | Electrocardiography |
| HMM | Hidden Markov Model |
| HR | Heart Rate |
| MFCC | Mel-Frequency Cepstral Coefficients |
| MRI | Medical Resonance Imaging |
| PCG | Phonocardiogram |
| S1 | Heartbeat associated with the onset of systole |
| S2 | Heartbeat associated with the onset of diastole |
| SNR | Signal-to-Noise Ratio |
| STFT | Short-Time Fourier Transform |
| SVM | Support Vector Machine |
| WHO | World Health Organization |